\documentclass[preprint,times]{aastex}
\usepackage{natbib}

\shorttitle{Angular diameter of Mira and semi-regular variables}
\shortauthors{Mondal, S. and Chandrasekhar, T.}

\begin{document}
 
\title{Angular diameter measurements of evolved variables by lunar occultations
at 2.2 and 3.8 $\mu$m}

\author{Soumen Mondal and T. Chandrasekhar}  
\affil{ Physical Research Laboratory, Ahmedabad-380 009, INDIA}
\email{soumen@prl.ernet.in; chandra@prl.ernet.in}

\begin{abstract}

We report the angular diameters  of two Mira  variables (U Ari and Z Sco),
three semi-regular and irregular variables  (SW Vir, $\eta$ Gem  and $\mu$ Gem)
and a supergiant semi-regular variable (TV Gem) by lunar occultation
observations in the near-infrared broad K-band (2.2 $\mu$m). Lunar occultations
of $\eta$ Gem  and $\mu$ Gem were also observed for the first time
simultaneously in both K and L$^\prime$ bands yielding angular diameters at 2.2
and 3.8 $\mu$m. Effective temperatures and linear radii are also derived for
all the observed sources and compared with earlier measurements. The mode of
pulsation of both Mira and SR sources in our sample is discussed.    

\end{abstract}


\section{Introduction} 

Asymptotic giant branch (AGB) stars are in the last stage of stellar evolution
before turning into planetary nebulae and are generally surrounded by  
circumstellar matter due to their large  mass-loss rates   ($\sim$ 10$^{-6}$ 
M$_\odot$ yr$^{-1}$). AGB stars include classical Mira variables (visual
amplitudes $>$ 2.5 mag ; periods, 100 - 1000 days), semi-regular variables SRa
(visual amplitude $<$ 2.5 mag ; periods 35 - 1200 days), semi-regular 
variables SRb (amplitude $<$ 2.5 mag ; with poorly defined periods), and
irregular variables Lb type (amplitude is small, no definite periods), as well
as supergiant semi-regular variables SRc type \citep{smith02}. The evolutionary
connection between Mira and SR group is not clear, but SR variables are often
considered to be Mira progenitors \citep{bed98a}. 

Multiwavelength measurements of angular sizes of Miras and Semi-Regular
variables at different phases of their pulsation cycle provide a direct means
to understand their atmospheric extension and pulsation properties. The high
mass-loss and relatively low surface temperature of the evolved stars provide a
habitable zone for several molecules like TiO, VO, H$_2$O \& CO etc. in their
extended atmospheres. Large opacities of these atmospheric molecules in some
particular bands mask the dominant photospheric continuum radiation. The
emergent observed radiation is thus contaminated by radiation from the
relatively cold atmospheric layers. Consequently photospheric size measurements
are effected in the different filter bands which has been known for sometime
\citep[e.g.][]{lab77, qui93, han95}.   

Recently there have been many high quality interferometric measurements of
angular diameters of Miras at near infrared wavelengths \citep[e.g.][]{menn02,
wruff04, per04, millan05, fedele05}. \citet{menn02} found that L$^\prime$-band
diameter of several oxygen rich Miras were much larger (25\% to 100\%) than
those measured in the broad K-band and proposed as an explanation a simple
empirical model of a central stellar disk surrounded by a optically thin
gaseous shell.  Observed variation  of angular size with wavelength can be
interpreted in terms of transparency of the optically thin shell varying with
wavelength. \citet{per04}  have observed several Miras in narrow bands around
2.2 $\mu$m and find systematically larger diameters in bands contaminated by
water vapor or CO. \citet{millan05} report a systematic increase of angular
size with wavelength ($\sim$ 25\%) from J to H to K$^\prime$ from a study of 23
Miras involving simultaneous measurements in JHK$^\prime$. The 11 $\mu$m
interferometric observations of three Mira variables \citep{weiner03a,
weiner03b} showed the diameters at 11 $\mu$m larger by a factor of $\sim$2 than
those measured in the K-band. The increase of apparent diameter from the
near-infrared toward longer wavelengths seems to be a common phenomenon in Mira
and late M-type semiregular variables. The dispersion in angular sizes from
near-infrared to mid-infrared band are well represented by the modeling of
interferometric data by inclusion of water shell surrounding the Mira variables
\citep{weiner04, ohnaka04, Schuller04}. ISO and ground-based spectroscopy
observations also point  to the warm molecular shell of H$_2$O surrounding the
Mira variables \citep{hin79, tsuji97, yam99, mat02, tej03b}. Several
theoretical models have been developed to understand the dynamic atmosphere of
Mira and non-Mira M stars \citep{bes96, hof98, hof98a, hofner98, hofner03,
woitke99}. Recent results of interferometric observations have been predicted
well by considering molecular contamination effects on theoretical models
\citep{jacob02, tej03a}. 

The question of pulsation modes for Mira stars is a complicated one.
Theoretical considerations generally suggest a fundamental mode as it is
difficult to reproduce in the first overtone, the large velocity amplitude
encountered in Miras \citep{bes96}. The quantity commonly used in stellar
pulsation modelling is the Rosseland radius of the hypothetical parent star of
the Mira variable which does not pulsate.It is not an observed quantity  but
related to the intensity distribution on the stellar disk \citep[][and
references therein]{scholz03}. It is also a phase and possibly cycle dependent
value. \citet{jacob02} showed that transforming a measured diameter into a
Rosseland value may be difficult or impossible due to molecular contamination
in even standard near continuum bandpasses. The variability of diameter with
Mira phase and cycle as well as wavelength further complicates the issue.
Angular diameter that are reported from observed data are obtained by fitting
visibilities (from interferometric data) or occultation lightcurves (from lunar
occultation) to a well defined artificial center to limb variation (CLV) like a
uniform disk (UD), a fully darkened disk (FDD) or a Gaussian intensity
distribution. It is difficult to obtain the actual CLV from the observed data.
Another difficulty in determining pulsation modes is the large uncertainty in
distance measurements required to convert angular to linear diameters. While
earlier work \citep{han95, vanl97, tej99, van02} suggested first overtone
pulsations in some Miras, recent results based on interferometric diameter
measurements \citep{per04, wruff04, fedele05} which take into into account
molecular contamination in the band passes \citep{jacob02, menn02, tej03a,
ire04a, ire04b, ire04c, ohnaka04} point to a fundamental mode of pulsation in
Miras.

In this paper we present new angular size measurements of six evolved variables
by lunar occultation (LO) in the near-infrared. These include two Miras (U Ari
and Z Sco) and four semiregulars (SW Vir, $\eta$ Gem, $\mu$ Gem and TV Gem)
including the supergiant TV Gem.  Our occultation results from another Mira U
Ori are also included for comparison although it's angular diameter has been
reported earlier \citep{mon04}. We report the first simultaneous angular
diameter measurements at 2.2 and 3.8 $\mu$m for two SR variables ($\eta$ Gem
and $\mu$ Gem). $\eta$ Gem underwent two LO events and in both cases
simultaneous K and L$^\prime$ angular diameters could be determined. Uniform
disk angular diameters are derived from our occultation data. Bolometric fluxes
are estimated from photometry and effective temperature are calculated.
Distances estimated to the sources are considered and linear radii are derived.
The position of these evolved variables is plotted in the period-linear radius
diagram and their mode of pulsation discussed.

\section{Observations and data analysis}  

Three  sources (U Ari, $\eta$ Gem and $\mu$ Gem) were observed simultaneously
in broad K (2.2/0.40 $\mu$m) and L$^\prime$ (3.8/0.60 $\mu$m)  filter-bands
while Z Sco, SW Vir, TV Gem and U Ori occultations were recorded in the K-band
only. All observations were made using two channel high speed photometer
installed on the 1.2m telescope at Mount Abu, India.  The details of the
instrument can be found elsewhere \citep{mon99, mon02}. The sampling time of
the light curves  was 2 milliseconds (ms) except SW Vir which was sampled at 1
ms.  $\eta$ Gem was observed twice during an interval of 2 months.  All events
expect Z Sco were recorded under good sky conditions, Z Sco event was recorded
through thin clouds. The details of occultation events are listed in Table 1.

The observed light curves containing modified Fresnel diffraction  fringes are
modeled to get the uniform disk angular diameter of the stellar sources. The
model fitting of the lunar occultation light curve involves the
$\chi^2$-minimization technique to obtain the best estimation of the five  
parameters : (i) the geometric time of occultation; (ii) the stellar signal;
(iii) the sky-background; (iv) the velocity component of the moon in  the
direction of occultation and (v) the uniform disk   angular diameter. The
analysis procedures are based on the standard non-linear least- square (NLS)
method introduced by \citet{nath70}. The point-source Fresnel diffraction
pattern modulated by the finite spectral bandwidth of the system, the finite
telescope aperture, the instrument time response and the extended angular size
of the source are taken into consideration for fitting of the above-mentioned
parameters. The resolution limit of the LO technique experimentally determined
by studying occultations of a number of bright point sources, is $\sim$ 2 mas
\citep{chand99}.

\section{Results and Discussions}

Individual source parameters are listed in Table 2. Mass-loss rates and
outflowing velocities derived generally from CO and SiO line measurements are
also listed in Table 2. The difficulty of obtaining good distance estimate is
discussed later in section 3.3 and adopted distances to sources are given in
Table 2. 

\subsection{Angular diameter measurements}

\subsubsection{U Ari}

{\bf{U Ari}} is an oxygen-rich Mira variable of period 371 days with a spectral
type M4-9.5 IIIe \citep{keen89}.

Lunar occultation of U Ari was observed  simultaneously in the K and L$^\prime$
bands  close to minimum phase ( phase 0.57).  A good occultation trace has been
recorded in K-band (Fig. 1).  The signal to noise ratio (S/N) limited by
atmospheric scintillation noise in K-band is about 50. The L$^\prime$-band
profile is noisy and has not been considered for analysis.      

We obtain a UD angular diameter of 7.3$\pm$0.3 mas in the K-band. The UD model
fit to the K-band light curve is shown in Figure 1. An earlier LO measurement
in the H-band gave a UD diameter of 6.11$\pm$0.34 mas at phase 0.49
\citep{rid79}. There is a significant difference of about 20\% in the two UD
values which have been measured at nearly the same phase near the minimum. From
recent simultaneous size measurements of 23 Miras at JHK$^\prime$-bands, the
larger size in K-band compared to H-band  appears to be a common phenomenon in
Mira stars \citep{millan05}. The apparent size variations from the near to
mid-infrared wavelengths are reasonably well modeled  by inclusion of warm
(1500-2000 K) H$_2$O shell within a few stellar radii \citep{menn02, jacob02,
tej03a, weiner04, ohnaka04, per04}. Furthermore, it is to be noted that as both
observing (K and H-band) phases are close to  the minimum, the molecular
contaminations are expected to be more prominent compared to the maximum phase
and also evident from spectroscopy \citep{tej03b}. Hence our simple fit of UD
diameter may not represent the true continuum size near the minimum of the Mira
phase. In case of interferometric observations, the observed visibilities are
compared successfully to the theoretical models of visibilities with inclusion
of a thin  surrounding H$_2$O shell \citep{per04}. However, in case of LO such
modeling efforts have not yet been carried out.

\subsubsection{Z Sco}

{\bf{Z Sco}} is an  oxygen-rich Mira variable of period 352 days with a spectral
type M4/5 IIIe. The maser lines of CO and SiO are not detected in the
circumstellar atmosphere of Z Sco \citep{young95, cho96} and  hence there is no
mass-loss estimate and outflow velocity of the source. No dust signature is
present in {\it{IRAS}} LRS spectra \citep{sloan98}.

The occultation of Z Sco was recorded in the K  band at phase 0.26. The sky
condition during observations was poor (thin passing clouds) but it was
possible to record the event. Fringe distortion is   evident in the occultation
trace (Figure 2) but nevertheless four fringes are recorded.  The light curve
is fitted with the UD model and a varying background using 5th order Legendre
polynomial shown in Figure 2.  

We first derive the UD  angular diameter of 3.8$\pm$1.0 mas. The error is large
due to poor quality of the light curve. The diameter of Z Sco has never been
measured before by a direct technique. Using the empirical relation of
\citet{van99}, based on (V-K) magnitude, we derive an angular diameter of
5.1$\pm$1.3 mas using the visual magnitude of 11.6  from the AAVSO database
\citep{math04}. We measure the K magnitude of the source to be 1.33$\pm$0.1 at
the time of our occultation observations.

\subsubsection{U Ori}

{\bf{U Ori}} is an oxygen-rich Mira variable of pulsation period 371 days with a
spectral type M6-M9.5 IIIe \citep{keen89}.

The occultation of U Ori was recorded in the K-band at phase 0.33. We obtain the
UD diameter of 11.90$\pm$0.30 mas. We had earlier studied this source from the
point of view of the spatial asymmetry. From the comparison of near simultaneous
lunar occultation observations on U Ori at the same wavelength (K-band) and at
different position angles (PA) (75$^o$ and 136$^o$), from two observations we
found the evidence of asymmetry in its atmosphere; the source appears to be
elongated at PA of about 70$^o$ \citep{mon04}. The asymmetric extension is also
found in OH maser observations on U Ori \citep{chap91}.      

{\subsubsection{SW Vir}

{\bf{SW Vir}} is an oxygen-rich semi-regular pulsating variable of SRb type
having a pulsation period of 150 days and spectral type of M7 III
\citep{leb99}.  \citet{kiss99} identified it as a triply periodic variable
(1700, 164 and 154 days). Here we have adopted the period of 150 days that is
widely accepted in the literature. 

The lunar occultation of SW Vir (M7 III) was recorded in the K-band only. The
S/N of data is limited by the atmospheric scintillation.  We derive the UD
diameter of 15.9$\pm$0.6 mas (Table 3).  The best-fit UD  model with observed
data-points are shown in Figure 5.   

Three angular diameter measurements of SW Vir was reported previously in the
near-IR region by LO and the reported values are in the range of 16.11 to 16.82
mas \citep{rid82, sch86}. Details of those measurements are listed in the Table
4. Recently using LBI the reported values of angular diameters are
16.24$\pm$0.06 mas in the K$^\prime$ band and 22.88$\pm$0.33 mas in the
L$^\prime$ band \citep{menn02}. The molecular contamination in SW Vir is
prominent from K and L$^\prime$ size measurements.  Our  measurement in the
broad K-band is consistent with previous measurements. From previously reported
measurements (in Table 4) we also note that phase-dependent variations are not
observed in SW Vir. 

\subsubsection{$\eta$ Gem} 

\bf{$\eta$ Gem}} is an oxygen-rich semi-regular variable (SRa) having a
spectral type of M2.5 III \citep{keen89}. The variability is classified as  SRa
with small visual amplitude of 0.75 mag. Two periods, 233 and 20 days, are
found from photometric observations \citep{percy01}. Most of the earlier
observations report  the longer period, 233 days.

Two lunar occultations of $\eta$ Gem (M2.5 III) have been observed  on 08
January 2001 and 4 March 2001 simultaneously in the K and L$^\prime$ bands. 
The  best-fit UD model curves with observed data points are shown in Figure 3.
We derive the UD angular diameters of 12.7$\pm$0.3 mas at K and 12.7$\pm$1.0
mas at L$^\prime$ for 08 Jan 2001 and 12.8$\pm$0.3 mas at K and 12.8$\pm$2.0
mas at L$^\prime$ for 04 Mar 2001 (Table 3).

We thus have two good sets of observations in both K and L$^\prime$ bands
separated by a two months interval. In this period a good K-band measurement of
angular diameter has been made \citep{ric03} which is in good agreement with
our value. From our two-epoch, two-wavelength observations we can conclude that
there is no detectable variation of angular size with phase in case of $\eta$
Gem.

Previous angular diameter measurements of $\eta$ Gem  are listed in Table 4.
The source has a well determined angular diameter from the wavelength range
0.55 to 2.2 $\mu$m. The optical (0.55 $\mu$m and 0.80 $\mu$m)  UD diameters are
11.43$\pm$0.55 and 10.91$\pm$0.11 mas respectively \citep{moz03}. At 0.712
$\mu$m (in strong TiO band) and at 0.754 $\mu$m (in adjacent continuum), UD
diameters are 11.75$\pm$0.27 mas and 10.70$\pm$0.15 mas respectively
\citep{qui93}. These optical diameters are not different from continuum
diameters and are slightly lower than our measured IR diameters.

$\eta$ Gem  is also identified as a spectroscopic binary. The spectral type of
the companion is  identified as G0 III with visual magnitude of 11.3 and the
separation of 0.9 to 1.08 arcsec from primary at position angle (PA) of 29$^o$
\citep{phil80, bai80}. The {\it{Hipparcos}} catalog shows a binary separation
of 1.7 arcsec at PA of 261$^o$ \citep{perry97}. The variation of separation in
several observations is attributed to the ellipticity of the orbit. No binarity
signature is detected in any of our LO light curves and it was also  undetected
from previous LO observations. The brightness ratio between the primary and
secondary component is estimated $\sim$ 1:1600 in the visual band. The K
magnitude of the companion would be $\sim$ 13 mag which is well below the 
limit of our detection.

\subsubsection{$\mu$ Gem}

{\bf{$\mu$ Gem}} is an oxygen-rich semi-regular variable of Lb type.The source
has a  spectral type of M3 III \citep {keen89}, a period of 27 days
\citep{percy01}.

The lunar occultation of $\mu$ Gem (M3 III) was also observed simultaneously in
the K and L$^\prime$-bands.  We derive the UD angular diameters of 13.7$\pm$0.5
and 14.8$\pm$1.0 mas in the K and  L$^\prime$-band respectively (Table 3). The
model-fit light curves along with observed data points are shown in Figure 4. 

There are thirteen observations of lunar occultation in the wavelength range
0.4 to 0.82 $\mu$m listed in the  catalog of  \citet{white87} and the mean UD
value in that wavelength range is 13.06$\pm$0.42 mas. Recently UD angular sizes
in the optical bands from interferometric observations are 13.98$\pm$0.14 mas
(at 0.80 $\mu$m) and 13.48$\pm$0.19 mas (at 0.55 $\mu$m) \citep{moz03}.   UD
sizes at TiO  absorption band ( 0.712 $\mu$m) and nearby continuum (0.754
$\mu$m) are 13.97$\pm$0.28 mas and 13.50$\pm$0.13 mas respectively
\citep{qui93}.  UD value at K-band  is 13.50$\pm$0.15 mas \citep{diben87}. 
Mira-like enlargement  (a factor of $\sim$2) at TiO band compared to the
adjacent continuum has not been noted in  $\mu$ Gem. Some of the previous
measurements are listed in the Table 4.    

Considering all available measurements including our own it appears that the UD
diameter of $\mu$ Gem has not shown any substantial variation from  optical to
near-IR over many years. Our UD angular diameters also show no significant
variation between K and  L$^\prime$ bands within the errors of measurements.

For non-Mira stars up to at least spectral type M4, it has been shown that
observed and model-predicted visibility curves do not differ significantly from
the UD profile \citep{wit01, wit04}. This is also realized from theoretical
models of non-Mira M giants \citep{hof98} as against the Mira model
\citep{hof98a}. In case of non-Mira stars ($\eta$ Gem, $\mu$ Gem and TV Gem but
not SW Vir) in our sample, UD size in the K-band may be close the true
continuum diameter consistent with previous observations (Table 4). Although
warm H$_2$O is detected in the spectra of K and early M giants \citep{tsuji01},
for $\eta$ Gem and $\mu$ Gem we find that there is no change between K and
L$^\prime$ diameters unlike in Miras or SW Vir (M7 III). Molecular layers if at
all present in the atmospheres of $\eta$ Gem and $\mu$ Gem must have column
densities too low to affect L band diameters unlike in Miras.

\subsubsection{TV Gem}

{\bf{TV Gem}} is an oxygen-rich supergiant semi-regular variable (SRc). The
spectral type is M1-0 Iab \citep{keen89}. The visual magnitude varies from 7.0
- 7.8 over the pulsation period of 182 days \citep{kuk69}. The distance we have
adopted here, is 1200$\pm$ 300 pc based on interstellar extinction towards Gem
OB1 association \citep{und84}. 

The lunar occultation of supergiant TV Gem  was recorded on 14 Nov 2000 in the
K-band under clear sky conditions. The S/N is high ($\sim$100) and is limited
by atmospheric scintillations. The UD model fit to the light curve with usual
five free parameters (in section 2) was not completely satisfactory.  A better
fit to the data is obtained by including a star plus shell model rather than a
single star model which is shown in the residuals of fits in lower panel of
Figure 6. The JHK photometric observations on 16 Nov of 2000  yielded the
magnitudes  2.31$\pm$0.05, 1.39$\pm$0.05 \& 1.16$\pm$0.06 in J, H and K
respectively.

Angular diameter measurements of TV Gem by lunar occultation have been reported
several times and are listed in Table 4. Earlier lunar occultations reported UD
values of 5.31$\pm$0.91 mas in the optical region \citep{rad84}, 4.9$\pm$0.3
mas in the K band \citep{rag97} and 4.46$\pm$0.07mas again in the K band
\citep{ric98}. From LO observations in the K-band \citet{rag97} had reported
the double shell structure of TV Gem like another supergiant $\alpha$ Ori
\citep{dan94}. The inner dust shell was estimated to be at 20$\pm$5 R$_*$. The
outer shell was estimated to be at $\sim$ 500R$_*$  based on LRS spectra   {\it
IRAS} and {\it IRAS} photometry (12, 25, 60 $\mu$m). They found the shell
contribution in the K-band to be $\sim$3\%. We measure the UD angular diameter
of the source 4.8$\pm$0.2 mas.  We  estimate the dust shell to be at 13$\pm$5
R$_*$. The shell contribution to the K-band flux is $\sim$5\% . These results
are consistent with earlier measurements by \citet{rag97}. We estimate the
effective temperature to be  3750$\pm$120 K again in good agreement with
earlier values of \citet{ric98}.

\subsection{Bolometric flux and effective temperatures }

The bolometric fluxes are estimated by  fitting a blackbody  curve to available
broad-band IR photometry measurements (JHKLM) compiled in  the Infrared
catalogue of \citet{gez99} and 12, 25 \& 60 $\mu$m {\it IRAS} PSC measurements
including our  JHK measurements in some cases. In some cases (U Ari, U Ori and
Z Sco) a two-temperature blackbody  is required to best fit all observed points
(1.25 -60 $\mu$m). Specifically an additional blackbody curve with cooler
temperature ($\sim$500 K) fits the excess in {\it IRAS} flux. Such fits are
shown in Figure 7.  For $\eta$ Gem a single temperature blackbody curve is
adequate for all observed fluxes (Fig. 7). The observed broad-band  photometric
magnitudes (JHKLL$^\prime$) are converted to flux densities using the zero
magnitude to flux density calibration established by \citet{bes98}. The
blackbody flux is normalized with the observed flux in the K-band.  In case of
two temperatures blackbody fit to the photometric data, the star flux is fitted
with relatively hotter temperature blackbody curve that is normalized with the
observed K-band flux. To fit the excess in the observed infrared fluxes with
additional relatively cooler temperature blackbody curve is further normalized
with the star continuum subtracted flux at 25 $\mu$m using the {\it IRAS} flux
at that wavelength.  By  numerically integrating the single (or resultant of
two temperatures) blackbody curve in the wavelength range 0.4 to 100 $\mu$m, 
the bolometric flux is calculated. No reddening corrections were applied to
estimate bolometric fluxes. These were deemed unnecessary, since typical
magnitude of the corrections  of our sample will be less than 0.05 mag in
K-band. For example, the largest visual extinction was found in U Ori, A$_v$
$\sim$ 0.25 mag \citep{white00a} and correspondingly A$_k$ $\sim$ 0.03 mag
using the wavelength-dependent extinction relation, A$_k$ = 0.11A$_v$
established by \citet{bes98}. Our own infrared JHK photometric measurements of
U Ari are used to estimate the bolometric flux at that particular phase (near
minimum) while others are taken from \citet{cat79} at a similar phase.  In case
of SW Vir and  TV Gem the bolometric fluxes were taken from the literature 
\citep{per98, rag97}. For other sources (U Ari, U Ori, Z Sco, $\mu$ Gem and
$\eta$ Gem ) we use our estimated bolometric fluxes from blackbody fits  to
calculate the effective temperature using the relation given below, 

\begin{equation} 
T_{eff} = 2341\times \left( \frac{F_{bol}}{\phi^2}
\right)^{1/4} 
\end{equation}

where the  bolometric flux F$_{bol}$ is in units of 10$^{-8}$ erg cm$^2$
sec$^{-1}$, the UD angular diameter $\phi$ is in milliarcsec and the effective
temperature T$_{eff}$ in K. The typical error in bolometric flux  is estimated
to be about 15\%. The effective temperature of the sources using our derived
K-band UD diameters are listed in Table 5. For Miras which are sources with
extended atmospheres , T$_{eff}$ refers to a specific layer. The characteristic
reference level seems to be near optical depth unity in the near-IR continuum
\citep{per04}. The effective temperature 2280$\pm$80 K of U Ari at phase 0.57
is substantially lower than the expected value of Mira stars ($\sim$3000 K)
while the value 2905$\pm$80 K of U Ori at phase 0.33 and 3120$\pm$420 K for Z
Sco at phase 0.26 are consistent. 

We estimated  the effective temperatures of 3450$\pm$125 K for $\eta$ Gem and
3675$\pm$140 K for $\mu$ Gem consistent with their spectral type. For SW Vir we
estimate the effective temperature to be 3060$\pm$130 K  consistent with
earlier measurements.

\subsection{ Linear radii and mode of pulsation }

From interferometric observations and theoretical models it appears that the
true continuum size estimate is difficult or impossible for Mira variables in
presence of molecular contamination effects and phase cycle effects.
\citet{fedele05} draw similar conclusions for R Leo. Other than wavelength and
phase effects, determination of the mode for pulsation in Mira stars from
period-radius relation  is also constrained by large  uncertainty in  the
distance to most of the sources as noted earlier in section 3 \citep{white00b,
vanl97}. Theoretical models \citep{bes96, hof98a} concluded on the fundamental
mode for these stars because it is difficult to produce the observed large
velocity amplitude \citep{hin97, scholz00}   from the first-overtone model. The
comparison of theoretical pulsation models with MACHO observations of
long-period variables in the LMC \citep{wood99}, pulsation velocities derived
from Doppler line profiles \citep{scholz00} also strongly indicate that Mira
stars are fundamental mode pulsators. Earlier observed UD diameters for Mira
stars at optical and near-infrared wavelengths without considering molecular
contamination pointed to the first-overtone mode \citep{han95, van02}.  More
recently considering molecular contamination on observed visibilities,
\citet{per04} arrived at a lower size compared to earlier UD-fit size of six
Miras and concluded that these are pulsating in fundamental mode.
\citet{wruff04} found that the interferometric visibilities curve of $o$ Cet
differs from UD profile and the data is better fitted with the fundamental
model.

SRs are separated from Mira stars by the shorter period and smaller amplitude,
and they often show evidence of multiple periods \citep{bed98, kiss99}. The
mode of pulsation of Galactic Miras are well studied while little attention has
been given to the pulsation studies on SRs. \citet{fea96} found that SRs in
globular clusters are pulsating in first overtone. From MACHO observations for
LMC red variables, \citet{wood99} concluded that SRs can be pulsating in the
1st, 2nd or 3rd overtone, or   the fundamental mode.  Comparing observational
and theoretical Q-values of 13 Galactic SRs, \citet{percy98} found that the
majority are pulsating in the 1st or 2nd overtone while some are pulsating in
the fundamental mode.

\citet{wood90} had suggested that in principle one combined equation for the
position of low mass AGB stars (both Miras and semi-regular variables)  can be
used for  comparing the observational results. The standard pulsation equation
is written as,

\begin{equation} 
Q = P\left(\frac{M}{M_\odot}\right)^{1/2}\left( \frac{R}{R_\odot}\right)^{-3/2} 
\end{equation}

where Q is a constant quantity (unit of days) that has distinct value for each
mode of pulsation. Here R is actually the Rosseland radius of the non-pulsating
star. Theoretical model predicted Q-values vary with period, mass and
luminosity \citep{fox82}.  Typically  the Q-value is $\approx$ 0.105
(fundamental) and $\approx$ 0.04 (1st overtone) estimated from the theoretical
models by \citet{fox82}. The Q-values used by \citet{percy98} for study of
galactic SR variables are 0.04 (1st overtone), 0.022 (2nd), 0.017(3rd) and
0.012 (4th) days from the models of \citet{xiong98}. We have estimated Q-values
for the samples considering mass of 1 M$_{\odot}$ (Table 6). Furthermore,
following \citet{ost86} we have considered the following expressions for
fundamental and overtone  modes respectively, 

\begin{equation}
\log P  = 1.86\log ( R/R_{\odot}) - 0.73 \log (M/M_{\odot}) -  1.92  
 \end{equation}

\begin{equation}
\log P  = 1.59\log (R/R_{\odot}) - 0.51 \log (M/M_{\odot}) - 1.60 
\end{equation} 

where P is the period in days.

Masses of Miras of moderate ($\le$400 days) period are reasonably well
constrained $\sim$ 1 M$_\odot$. \citet{wyatt83} estimated the main sequence
masses of 124 Miras considering available data on radial velocity measurements.
The Main-Sequence (MS) mass of progenitor of one Mira (U Ari) in our sample has
been determined by \citet{wyatt83} to be 1.3M$_\odot$. \citet{jura92} suggested
that the MS masses of Mira  progenitors are in the  range 0.8 - 2.0 M$_\odot$
for Miras having the period less than 400 days. Theoretical models of Mira
stars for masses of 1 and 1.2  M$_\odot$ have predicted that geometric
pulsation of continuum-forming layers is little affected by the mass difference
\citep{ire04b}. As semi-regular variables are progenitors of Miras, such mass
considerations may be also applicable for them. In this mode analysis we have
considered the mass range of 1.0 - 2.0 M$_\odot$.

The linear radius is obtained from our K-band UD angular diameters and the
adopted distance (Table 6) and superimposed on the theoretical model curves
(Fig. 8). The  errors shown on linear radii are mainly due to  errors on
distance estimates. 

There has been a great deal of discussion on {\it Hipparcos} parallax
measurements of Miras in the literature\citep{vanl97, white00b, knapp03}. The
revised {\it Hipparcos} parallaxes for U Ori and Z Sco have large errors (Table
6) and so we have not adopted them for distance estimation. {\it Hipparcos}
parallaxes are excellent database for $\mu$ Gem, $\eta$ Gem and SW Vir, and we
have adopted those parallaxes (Table 6). The adopted distances to all Miras in
our sample have been estimated  using the Period-Luminosity (PL) relationship
given in \citet{white00b} (M$_K$ = - 3.47 $\log$P + 0.84, M$_K$ is the absolute
K magnitude and P the period) and the reddening corrected mean apparent K
magnitude from \citet{white00b} (U Ori and Z Sco) and \citet{fea96} (U Ari).
The PL relations for Miras have been developed by several other authors
\citep{fea96, vanl97, knapp03} and those are consistent with the adopted
relation. Considering  uncertainties in apparent magnitudes (because of
variability amplitude) and the dispersion in PL relations, we have considered
errors in the Mira distances to be about 20\%.

From Figure 8 and the Q-value in Table 6, it appears that the Mira star U Ari
is pulsating in the 1st overtone mode from both K and H-band measurements.
However as noted in section 3.1.1 observations in both K and H bands were near
the minimum phase when molecular contamination effects are at a peak.
Converting the measured LO UD diameter directly to a linear diameter would tend
to overestimate the linear size. Till further measurements at other phases are
available the result of overtone pulsation in U Ari must be treated with
caution.

The Mira star Z Sco is probably a fundamental mode pulsator but precise angular
diameter determination was not possible because of noisy data. The mode of
pulsation for U Ori appears to be a borderline case. We note that similar
conclusions were drawn from earlier K-band interferometric observations of
\citet{van02}. The minimum diameter for U Ori  from  recent interferometric
observations in several narrow bands inside broad K-band \citep{per04} however
favours the fundamental mode. 

It is difficult to draw any conclusion on the SR variables SW Vir and $\mu$ Gem
from their position in Fig. 8. However, comparing  observational (Table 6) and
theoretical Q-values of \citet{percy01}, both SW Vir $\mu$ Gem may be the 
overtone candidates. The SRa $\eta$  Gem could be a fundamental mode pulsator
of low mass but it has a second period of 20 days which complicates the
issue.  

\section{Summary and conclusions}
 
Our UD angular diameter of U Ari in the K-band shows a substantially larger
value ($\sim$ 20\%) compared to H-band observed earlier at nearly the same
variability phase. Such an enhancement  is consistent with  a hot extended
molecular layer close to the photosphere as suggested by \citet{per04}.  

The supergiant TV Gem in the K band yields the angular size of 4.80$\pm$0.20
mas consistent with previous measurements. The  dust shell around TV Gem is
re-confirmed.  We measure the dust shell size to be 13$\pm$5 R$_*$. The
effective temperature derived is 3750$\pm$120 K, consistent with the earlier
value. 

We estimate linear radii of three Mira variables (U Ari, U Ori and Z Sco) from
the K-band lunar occultation uniform disk angular diameters and  distances
derived from the PL relation. Comparing theoretical Period-Radius plots (Fig.
8) we find that U Ari is a first overtone pulsator, Z Sco is  probably a
fundamental mode pulsator while U Ori is a borderline case between fundamental
and first overtone modes. A stronger conclusion regarding the pulsation mode of
Miras from occultation observations could  probably be reached by modeling
lunar occultation light curves for molecular contamination effects.

The two SR variables $\eta$ Gem and $\mu$ Gem clearly do not show any variation
in their angular diameter in K and L$^\prime$ bands unlike the Miras. SW Vir
and $\mu$ Gem appear to be candidates for overtone mode from their Q-values.
However as we are using occultation UD radii which are greater than Rosseland
radii, our Q-values may be less than the theoretical values. The case of 
$\eta$ Gem  is complicated by the presence of two periods in its optical light
curve. Further high angular resolution studies of SR variables are clearly
warranted.

\acknowledgments

The authors wish to thank the referee for many valuable suggestions for
improving the paper. This research has made use of the AFOEV and SIMBAD
databases, operated at CDS, Strasbourg, France. This work was supported by
Dept. of Space, Govt. of India.

\clearpage

\begin{figure}[p]  
\plotone{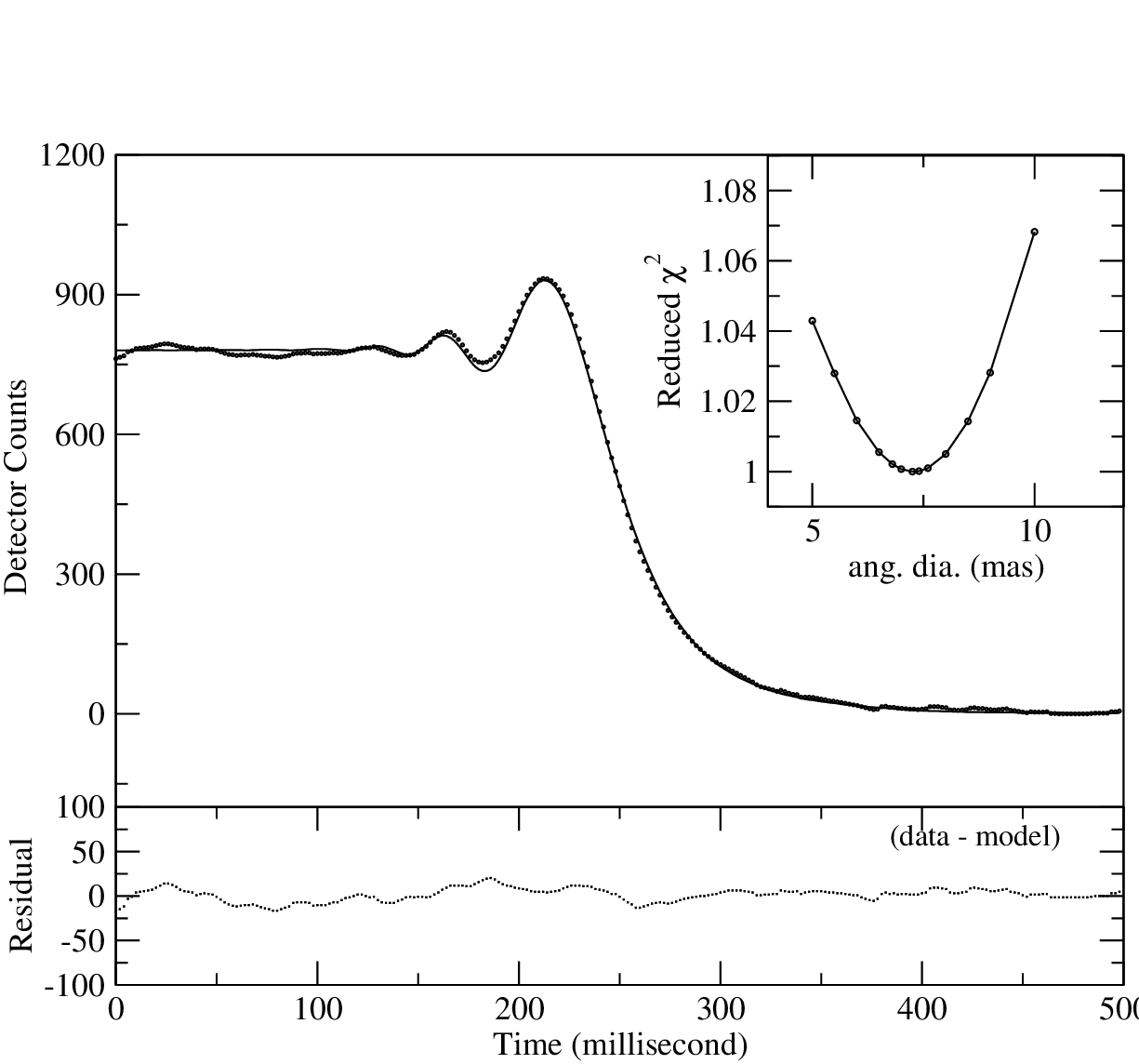}

\caption{Occultation light curves of U Ari  in the K-band:  model fit,
residuals of fit (lower panel) and convergence of fit (inset).  The dotted and
solid line is the observed data  and  model fit curve. The bestfit uniform disk
(UD) angular diameter is 7.3$\pm$0.3 mas.}  \end{figure}

\begin{figure}[p] 
\plotone{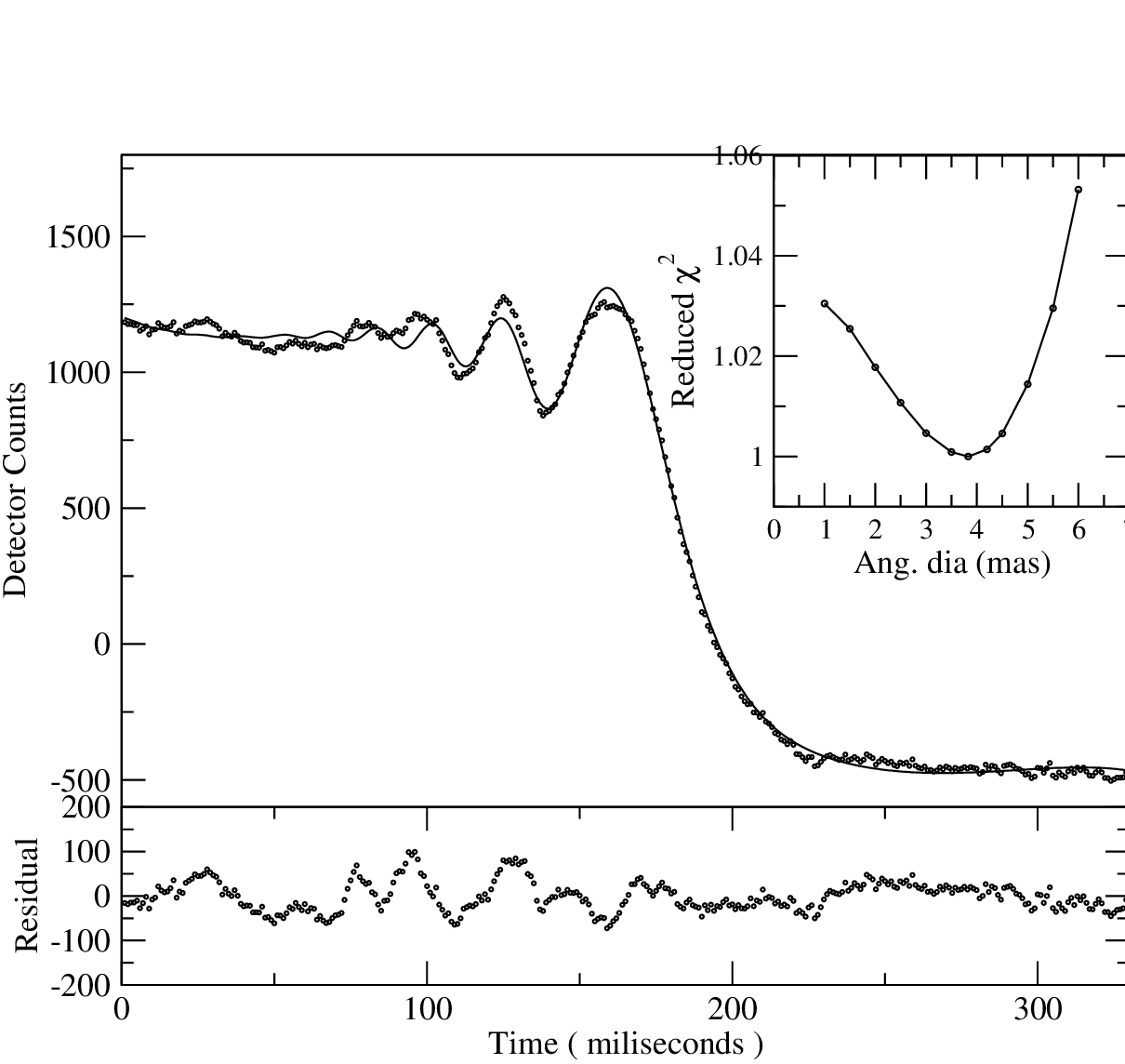}

\caption{Occultation light curves of Z Sco  in the K-band:  model fit,
residuals of fit (lower panel) and convergence of fit (inset).  The dotted and
solid line is the observed data  and  model fit curve. The bestfit uniform disk
(UD) angular diameter is 3.8$\pm$1.0 mas.}  \end{figure}

\begin{figure}[p] 
\plottwo{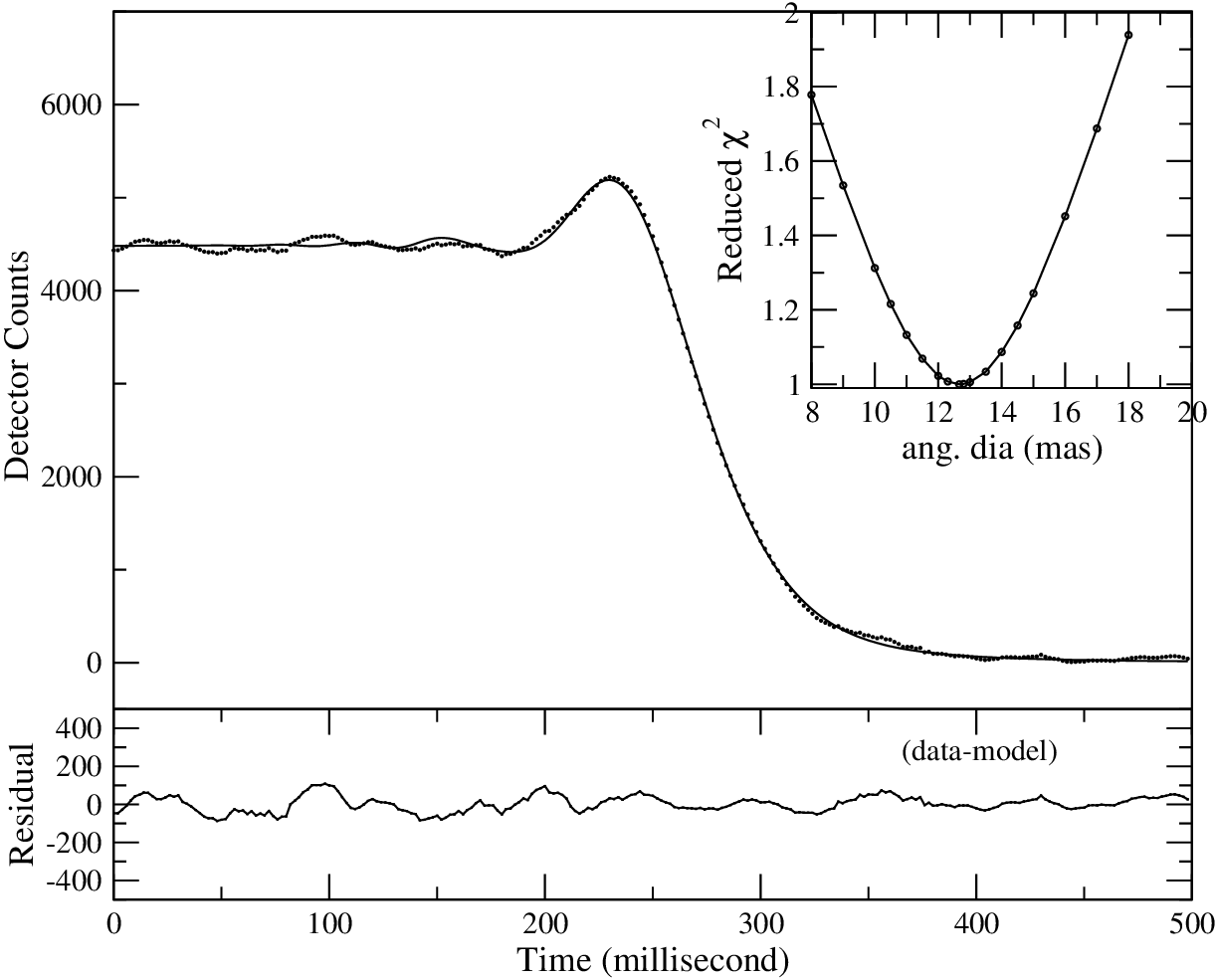}{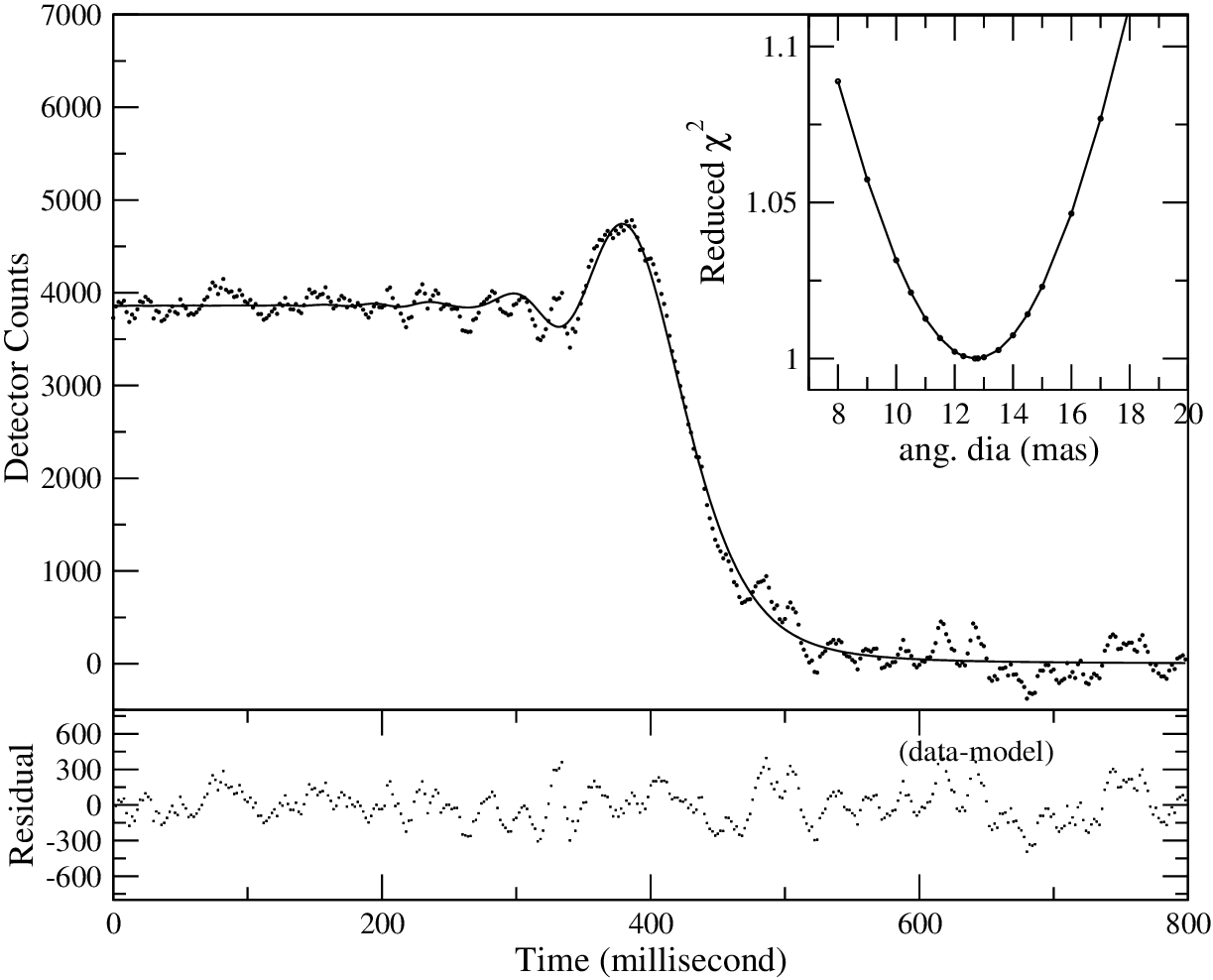} 
\vskip 0.5in 
\plottwo{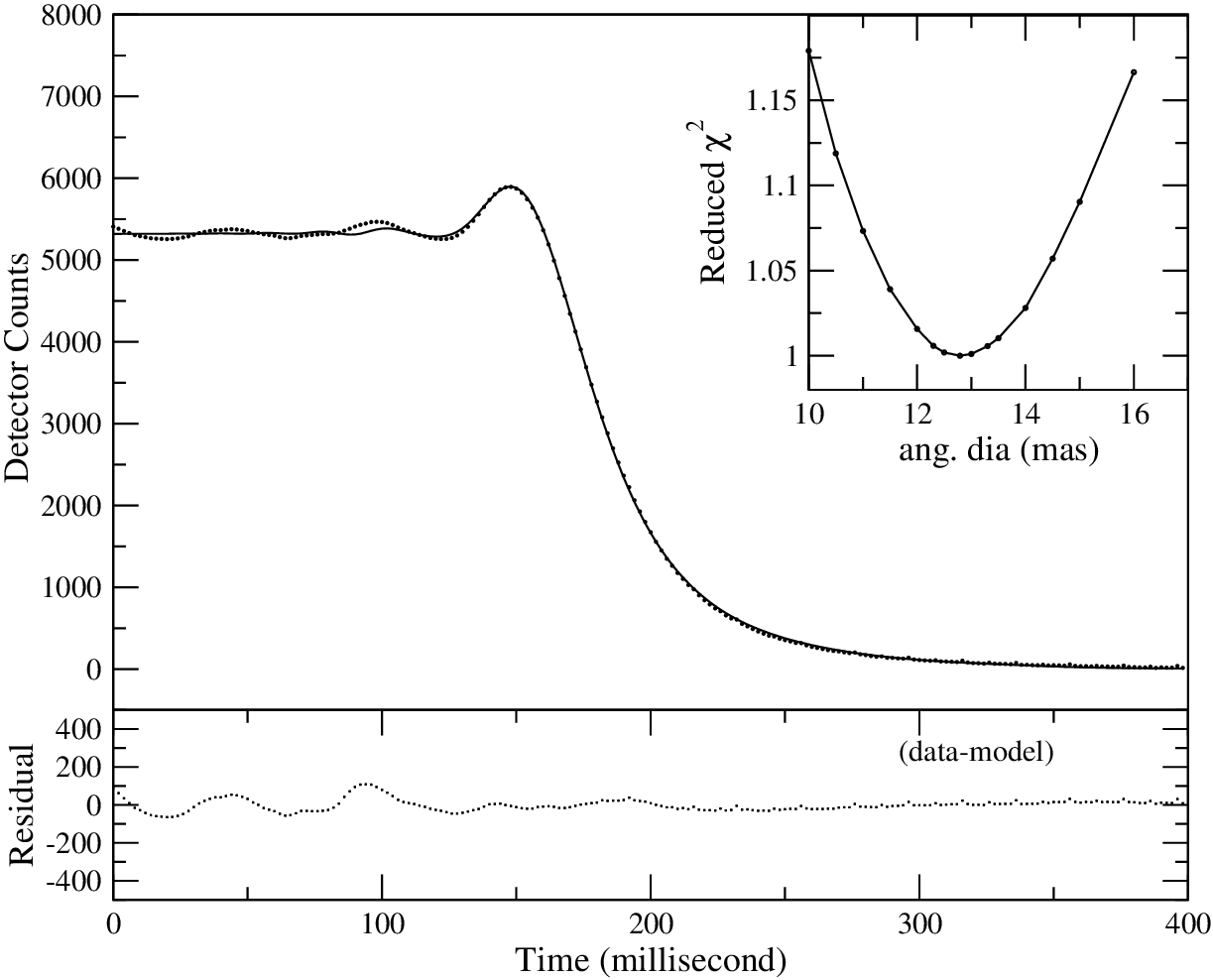}{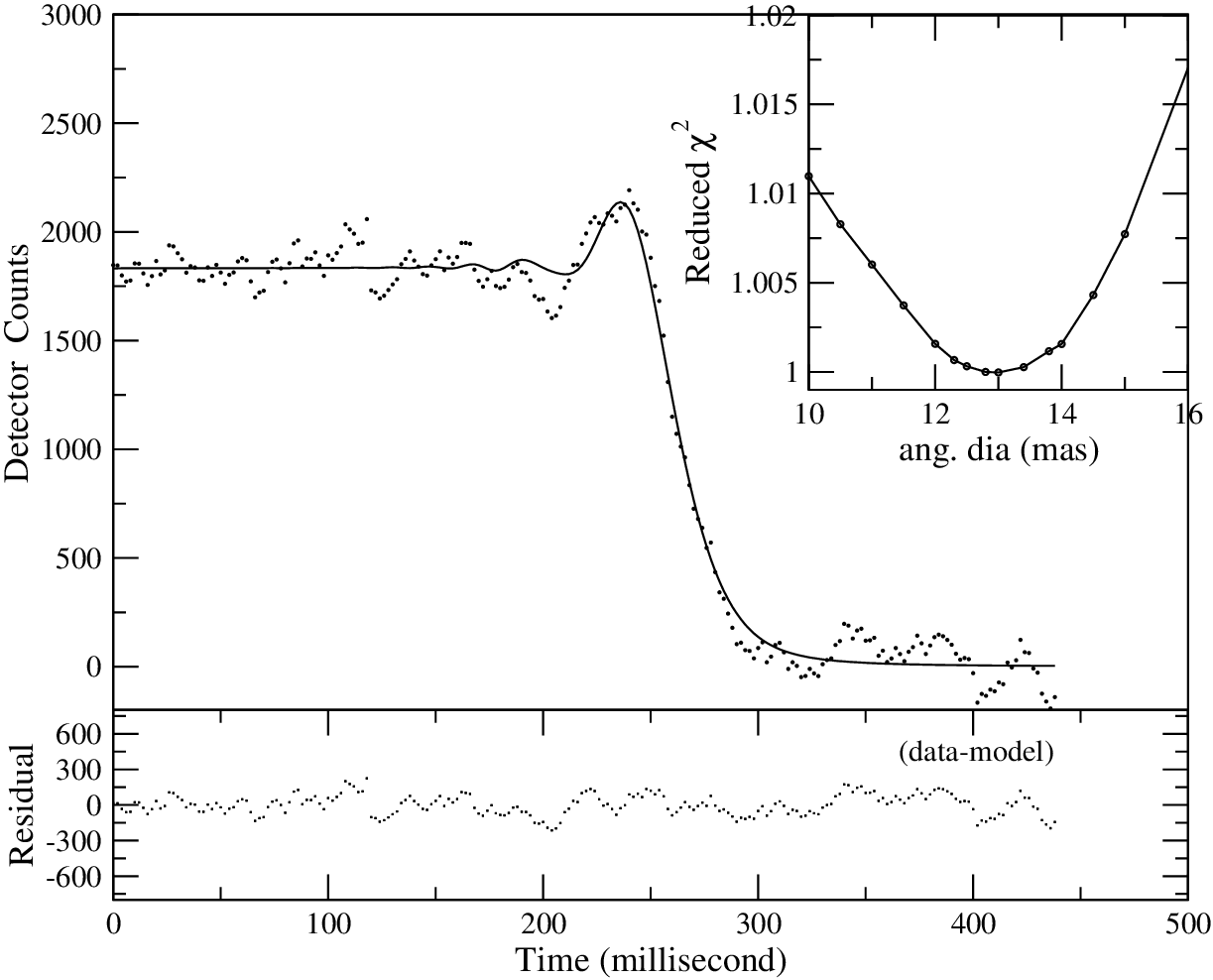}

\caption{Occultation light curves of $\eta$ Gem on 8 January 2001 (top)
and 4 March 2001 (bottom) and  : model fit, residuals of fit (lower panels) and
convergence of fit (insets) in the K-band (left)  and L$^\prime$-band (right).
The  dotted and solid line is the observed data  and  model fit curve. The
best-fit uniform disk diameters are  12.7$\pm$0.3 mas at K and 12.7$\pm$1.0 mas
at L$^\prime$ (8 Jan 2001) and 12.8$\pm$0.3 mas at K and 12.8$\pm$2.0 mas at
L$^\prime$ (4 Mar 2001) respectively.}   \end{figure}

\clearpage

\begin{figure}[p] 
\plottwo{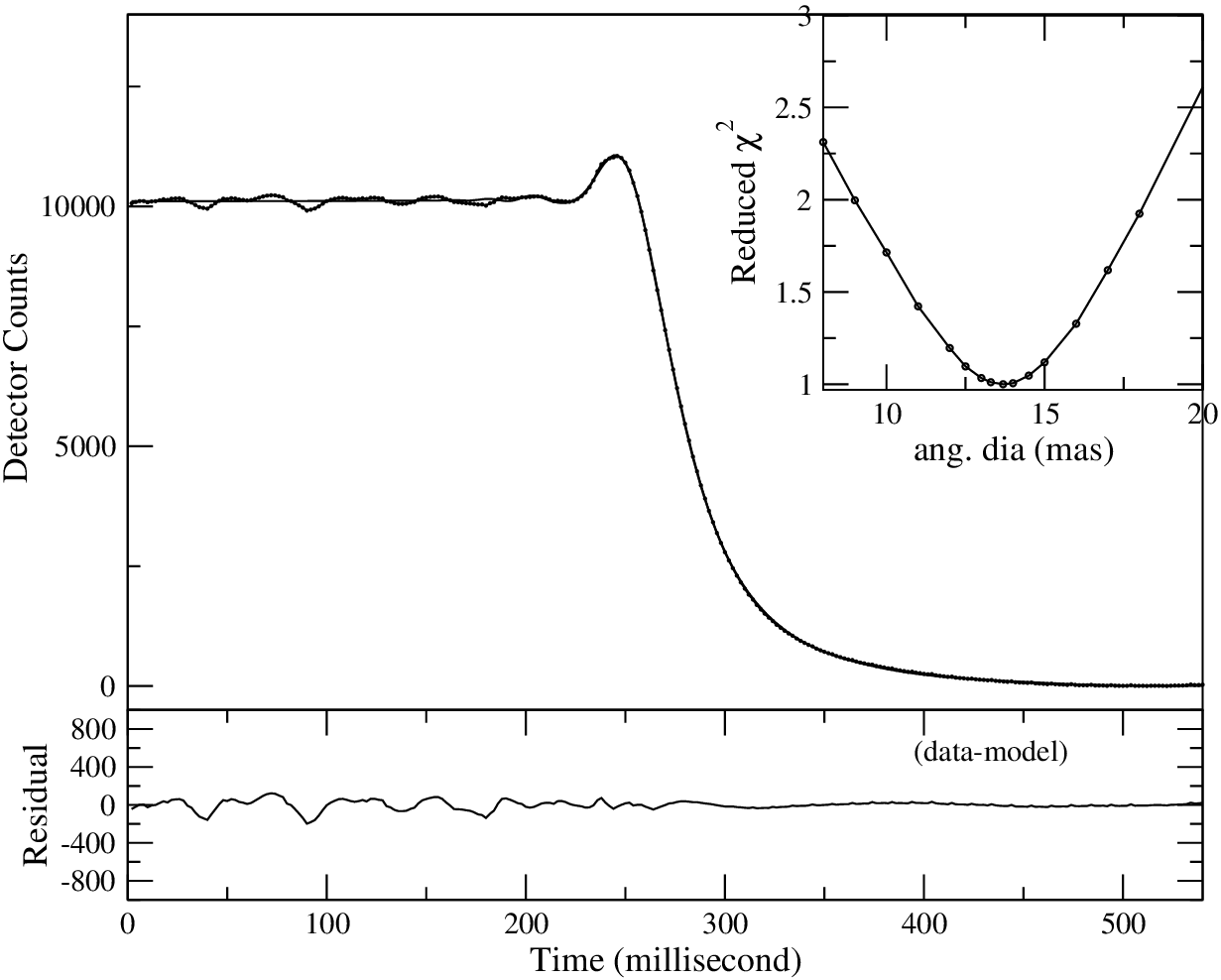}{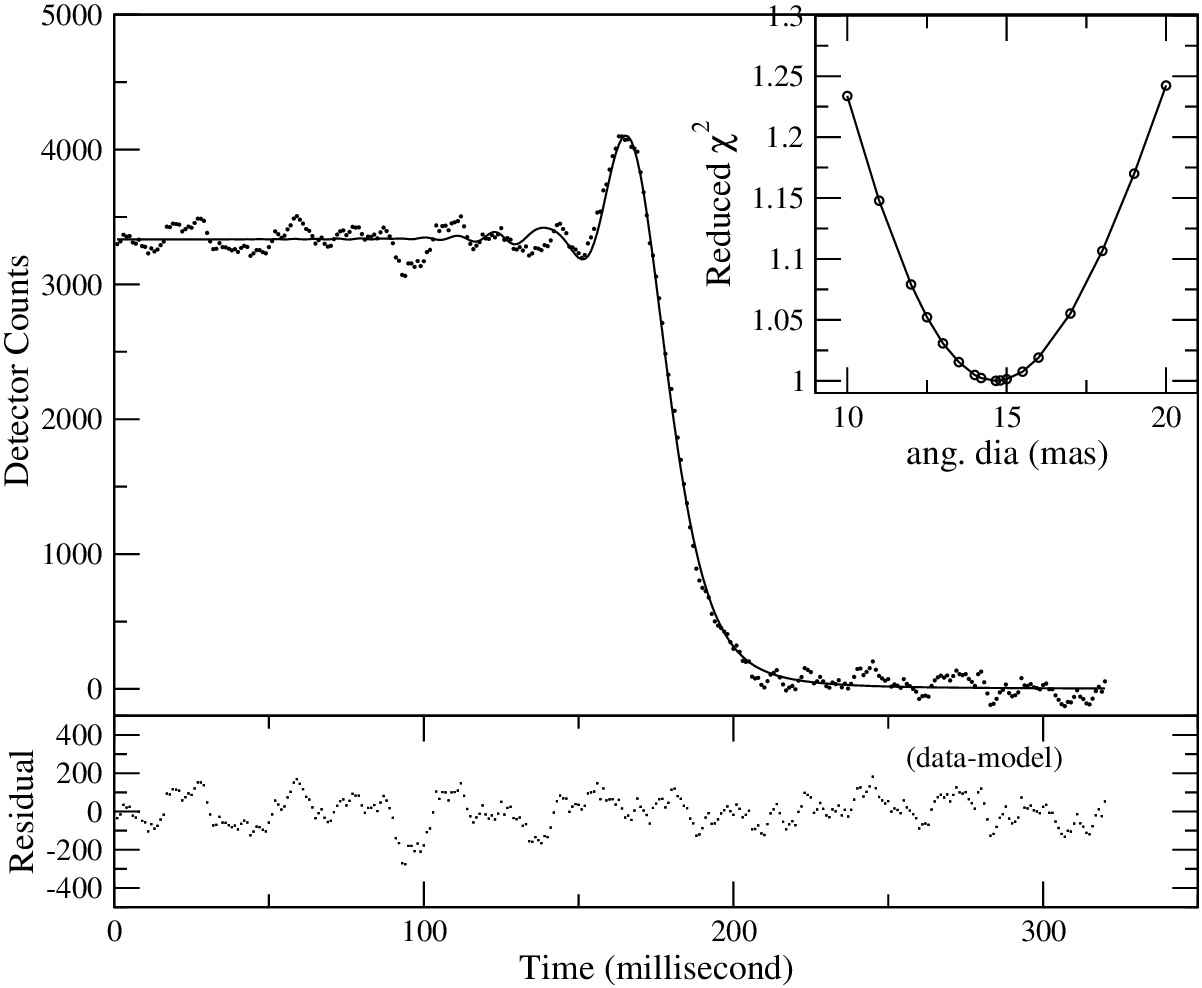}
\caption{\em  Occultation light curves of $\mu$ Gem in the K-band (left)  and
L$^\prime$-band (right) : model fit, residuals of fit (lower panels) and convergence of
fit (insets). The dotted and solid line is the observed data and  model fit
curve.   The best-fit uniform disk angular diameter is 13.7$\pm$0.5 mas at
K-band and 14.8$\pm$1.0 mas at L$^\prime$-band.}    \end{figure} 

\clearpage

\begin{figure}[p]
\plotone{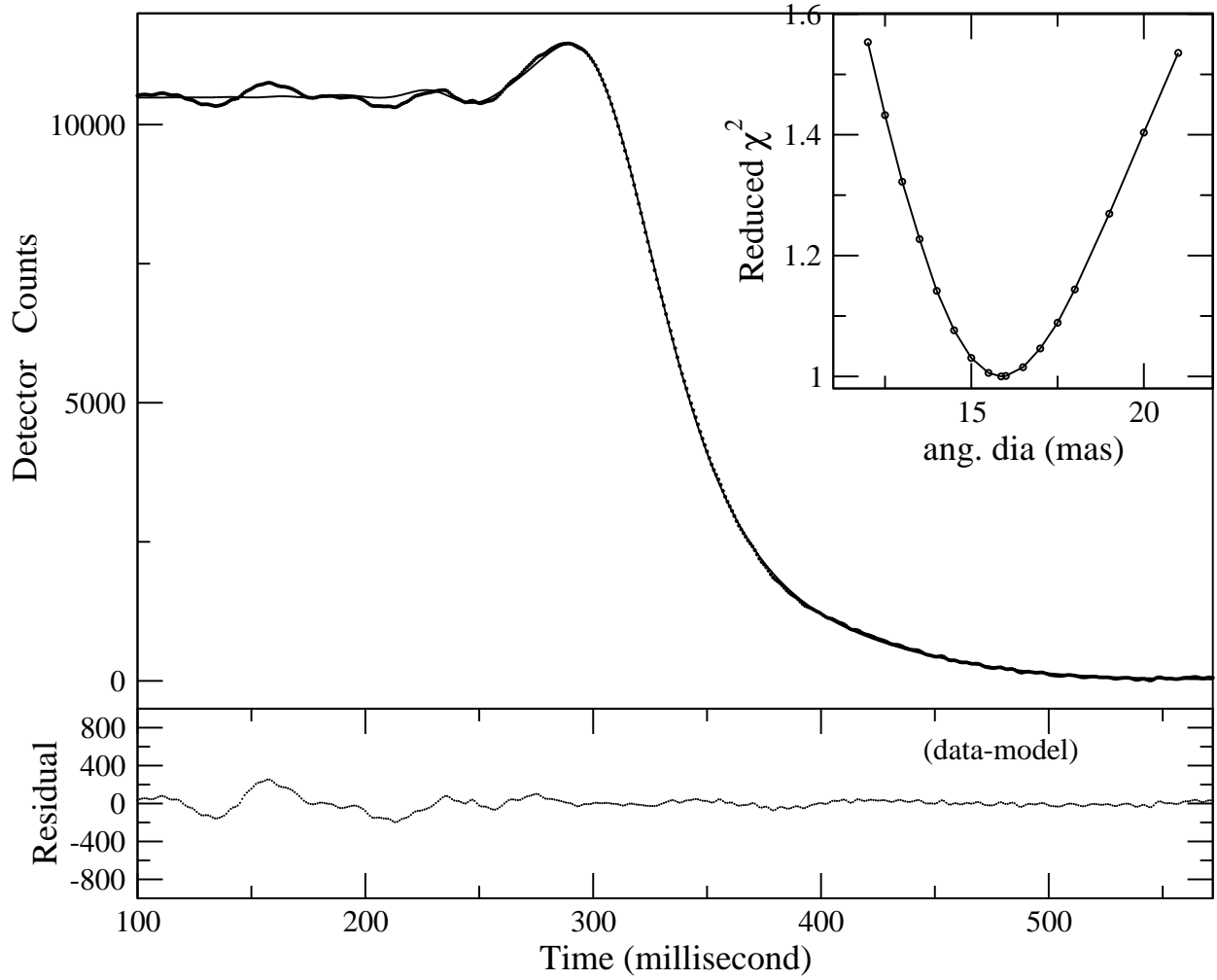} 
  
\caption{\em Occultation light curves of SW Vir  in the K-band: model fit,
residuals of fit (lower panel) and convergence of fit (inset).  The  dotted and
solid line is the observed data  and  model fit curve. The bestfit  UD angular
diameter is 15.9$\pm$0.6 mas} \end{figure} \clearpage

\begin{figure}[p]
\plotone{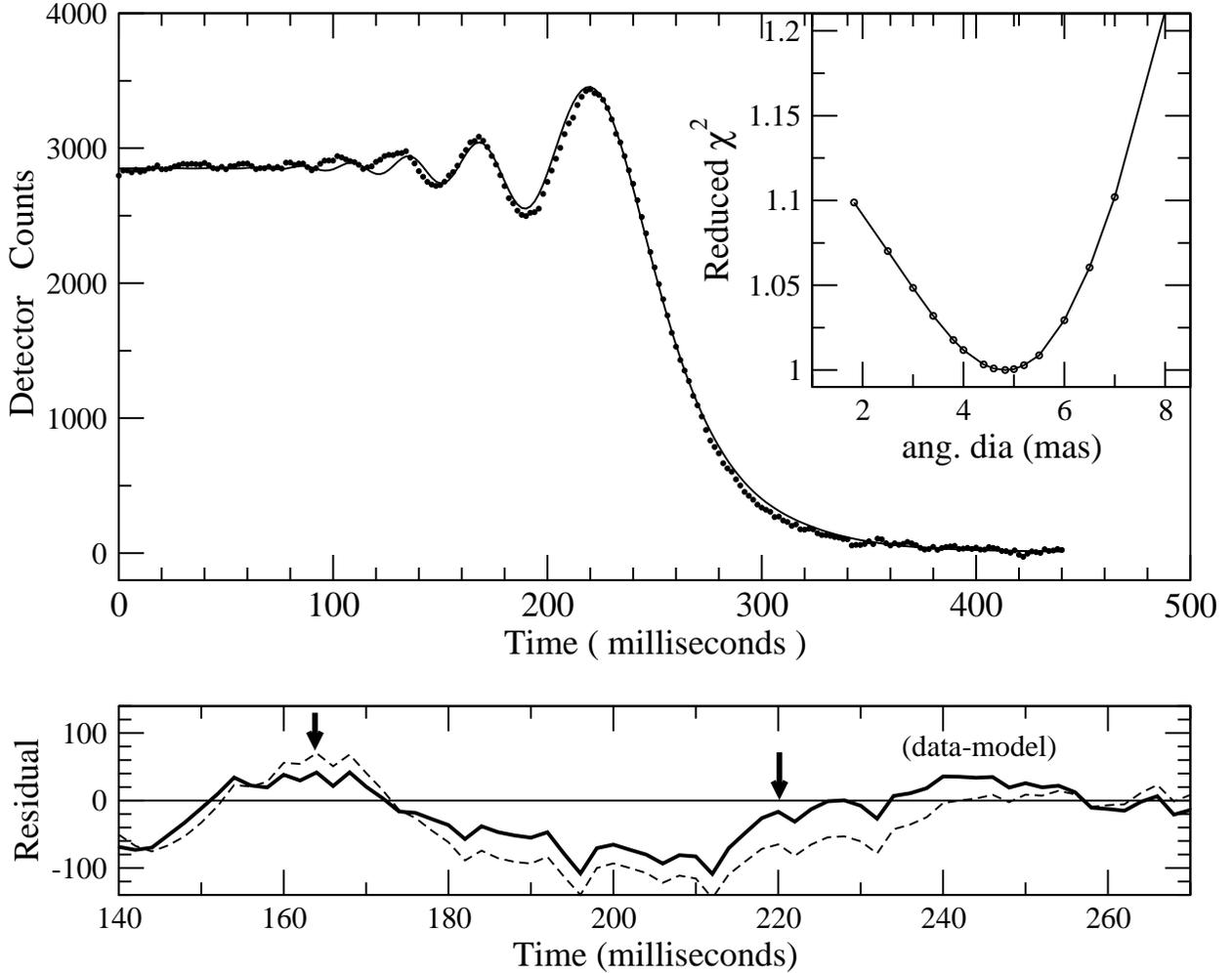}
   
\caption{\em Occultation light curves of  TV Gem  in the K-band : model fit and
convergence of fit (inset).  The  dotted and solid line is the observed data 
and  model fit curve. The best-fit  UD angular diameter is 4.8$\pm$0.2 mas.
Lower panel of the graph shows the enlarged residuals particularly on 1st and
2nd fringe positions which is marked by arrow for star+shell model fit (solid
line) and single star model fit (dotted line).} \end{figure} 
\clearpage

\begin{figure}[p]
\plottwo{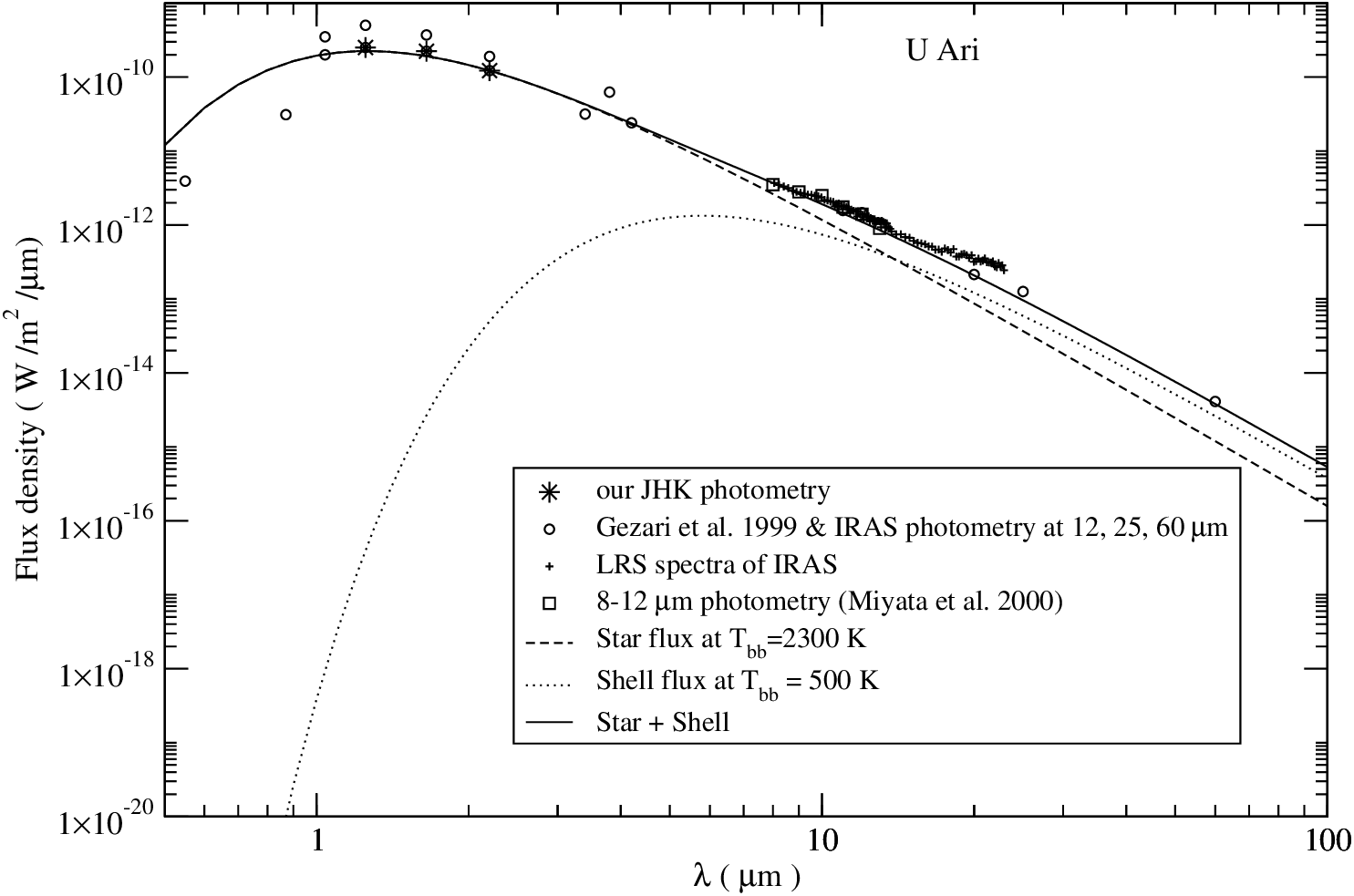}{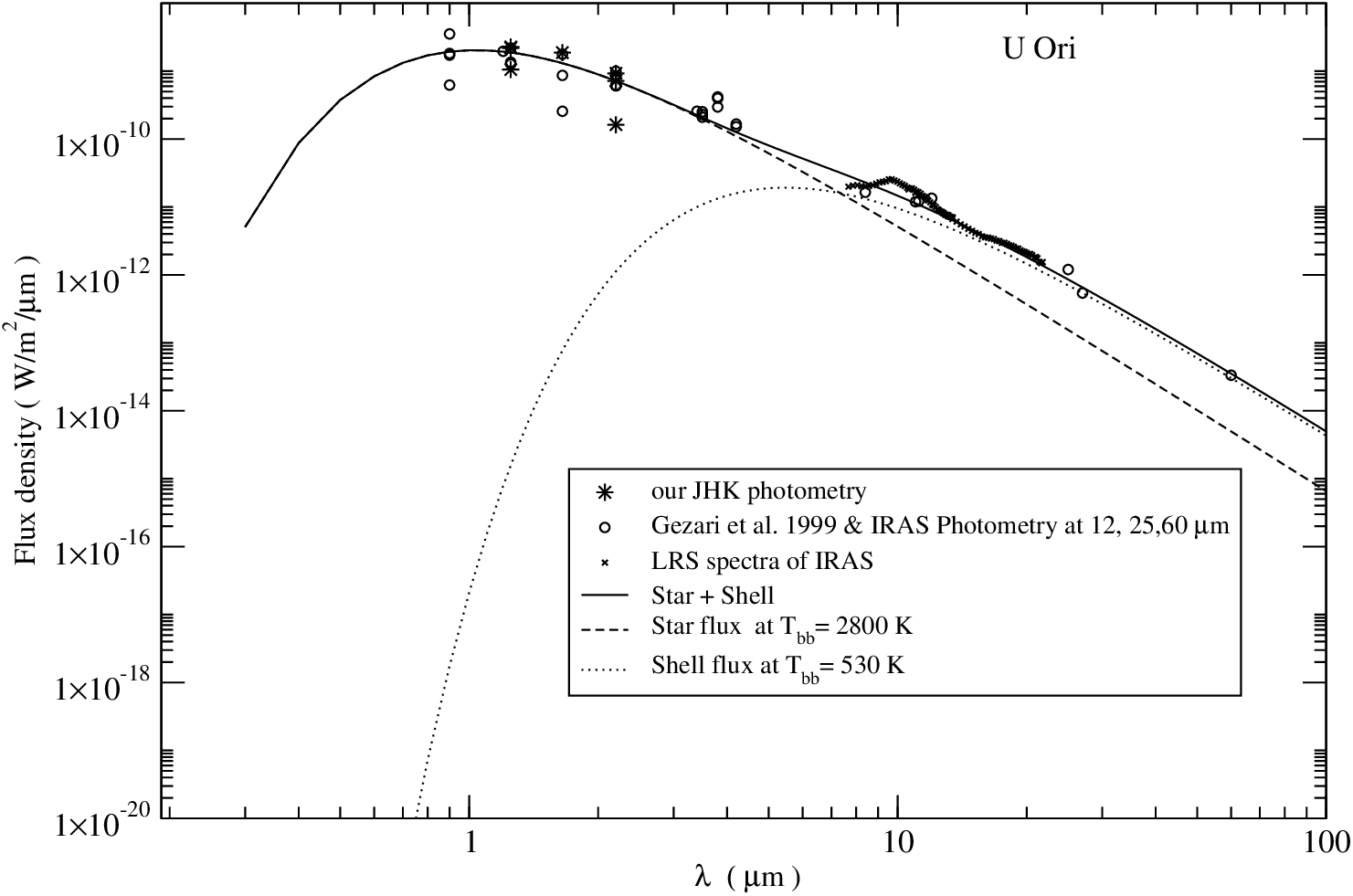} 
\vskip 0.5in 
\plottwo{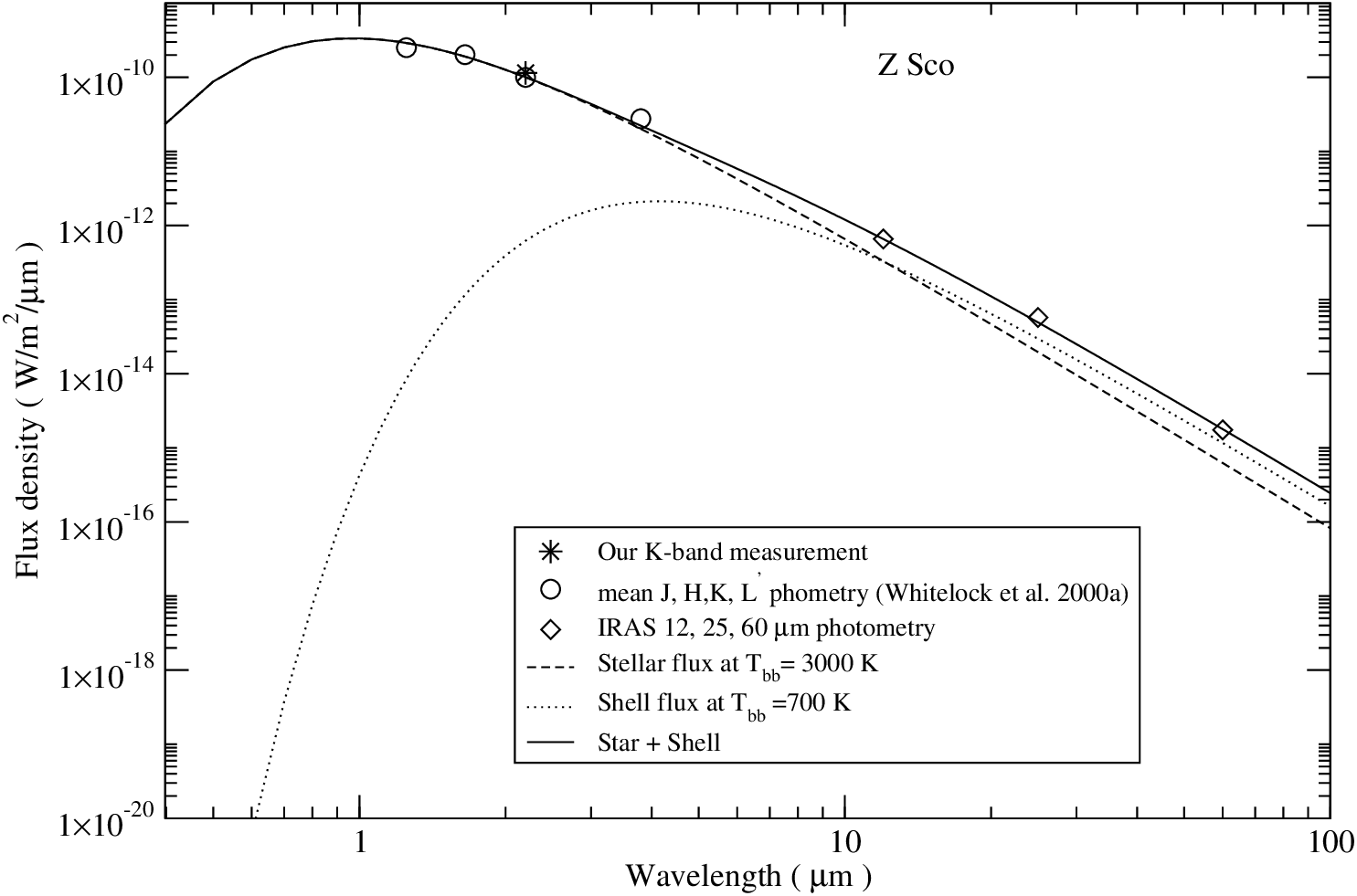}{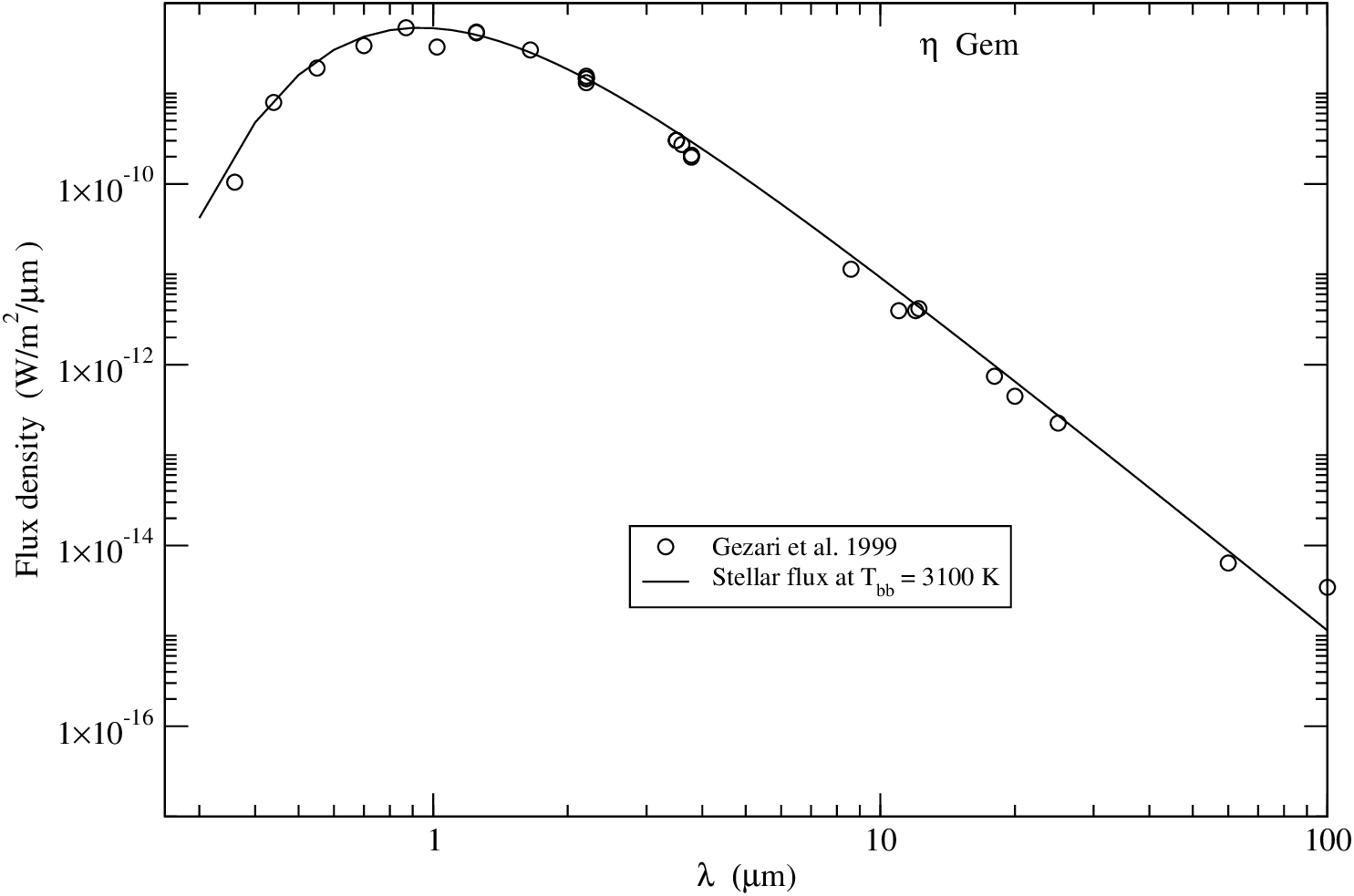}
\caption{\em The blackbody  fit to  photometry data available  in the
literature for U Ari, U Ori, Z Sco and $\eta$ Gem.} \end{figure}

\clearpage
  
\begin{figure}[p]
\plotone{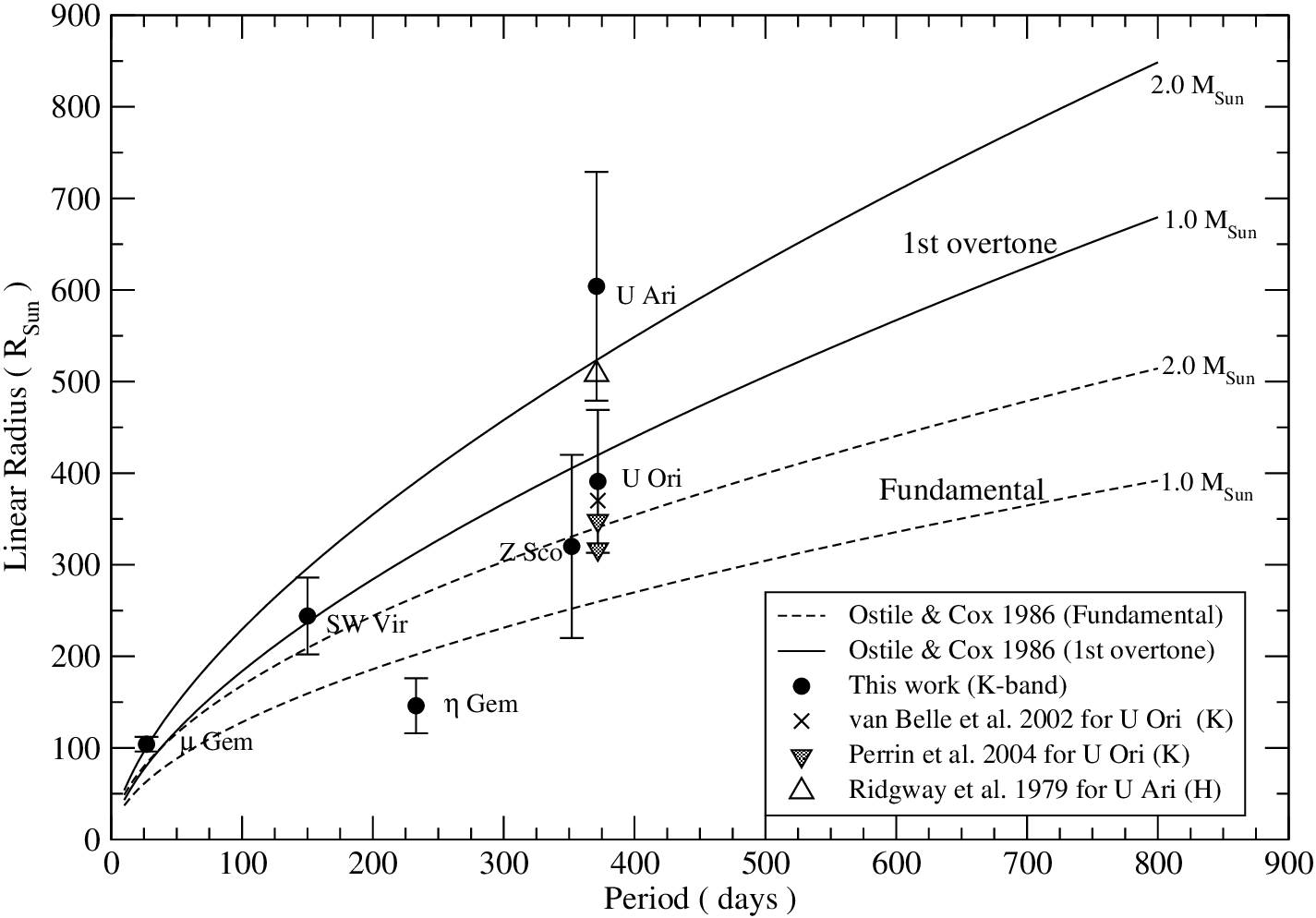}   
 
\caption{\em Linear radius vs. period plot for the Miras and semi-regular
variables . The filled circles represent our LO measurements in the K-band. The
fundamental and first-overtone curves in the mass range 1.0 M$_\odot$ - 2.0
M$_\odot$ are derived from the radius-period relation of eqn. 3 and eqn. 4.
Some of available linear radii for U Ori and U Ari in the literuare are
superimposed.}
\end{figure}

\begin{table}[p] 
\caption{\bf Log of Observations}
\vskip 0.2in
\begin{tabular}{ccccccc}  
\hline
\hline
Star (var.)  & IRC   & Date of & Filter($\mu$m)  &  PA &   v$_{comp}$ & event
type$^a$  \\  
     &       &   obser. &  $ (\lambda$/$\Delta\lambda$)  & (deg.)  & (km s$^{-1}$) & \\
\hline
U Ari (Mira) & +10040 &  19 Feb 2002 & 2.2/0.4  & 18  & 0.475& D   \\
       &        &                                 &  3.8/0.6 &   &    \\
Z Sco (Mira) & -20306 &  22 Mar 2003 &   2.2/0.4 &  155 & 0.651 &  R \\
       &        &          &      &               &       &   \\         
U Ori (Mira) & +20127 &  13 Mar 2000 &   2.2/0.4 &  136 & 0.66 &  D \\
       &        &          &      &               &       &   \\         
SW Vir (SRb)  & +00230 &   01 Jun 2001 &  2.2/0.4 & 74 & 0.504 & D \\
      &        &          &      &               &       &   \\       
$\eta$ Gem (SRa)  & +20139 & 08 Jan 2001 &  2.2/0.4 & 21 & 0.330& D \\
      &                &  &               3.8/0.6 &   &     \\
       &         &       04 Mar 2001 &  2.2/0.4 & 68 & 0.645& D \\
       &         &                       &   3.8/0.6 &   &     \\       
$\mu$  Gem (Lb)& +20144 &   04 Mar 2001   & 2.2/0.4 & 77 & 0.767& D \\
       &          &                        &   3.8/0.6 &   &      \\
TV Gem (SRc)   & +20134 &   14 Nov 2000 &  2.2/0.4 &  212 & 0.437 & R \\
   &        &               &            & 	&    &   \\
 
\hline
\end{tabular} 
\vskip 0.1in
$^a${\small D: Disappearance; R: Reappearance }\\
\end{table}

\begin{table*}[p]
\caption{\bf{Individual source parameters}} 
\vskip 0.2in
 
\begin{tabular}{lccccccc}
\hline
\hline  
Parameters &  U Ari & Z Sco & U Ori  & $\eta$ Gem &  $\mu$ Gem & SW Vir &  TV Gem \\
           &         &            &           &        & \\
\hline
Spectral type$^1$ & M4-9.5 IIIe & M4/5 IIIe  & M6.5-9 IIIe  & M3 III & M3 III & M7
III & M1-0 Iab \\
 &       &    &    &   &     &     &    \\ 
V mag  $^1$   & 7.2-15.2 & 8.7-13.4 & 5.3-12.6 & 3.20-3.90   & 3.20  & 8.2-9.4 &
8.7-9.5  \\
{\em(max-min)}   &       &    &    &   &     &     &    \\ 
K mag $^{2}$ & 1.59  & 1.48 & -0.49 & -1.49 & -1.89& -1.87&0.99 \\
   &       &    &    &   &     &     &    \\ 
L$^\prime$ mag $^{2}$ & 0.28 & 1.50 & -1.41& -1.59& -2.01& -2.28 & 0.62 \\
   &       &    &    &   &     &     &    \\ 
Variability type     & Mira & Mira & Mira & SRa  &  Lb  & SRb & SRc  \\
 &       &    &    &   &     &     &    \\ 
Period (days)$^1$ & 371 & 352 & 372 & 233 & 27$^{10}$ & 150  & 182  \\
     &       &    &    &   &     &     &    \\ 
Distance (pc) & 776$\pm$155$^3$ & 778$\pm$145$^3$ & 306$\pm$61$^3$ & 107$\pm$22$^4$ & 
71$\pm$5$^4$ & 143$\pm$24$^4$ & 1200$\pm$300$^5$\\
 &       &    &    &   &     &     &    \\   
Luminosity$^{11}$   & 8.4$\pm$3.5 & 8.3 $\pm$3.3 & 9.7$\pm$4.0 & 2.3$\pm$1.0 
&1.8$\pm$0.36 & 4.7$\pm$1.7 & 68.5$\pm$34.9  \\
{\em(10$^3$ L$_{\odot}$)}    &       &    &    &   &     &     &    \\  
Mass loss  & 57$^6$ & - & 29$^7$  &1.4$^8$ & 0.44$^8$  & 17$^7$ & 200$^9$  \\
{\em (10$^{-8}$ M$_\odot$ yr$^{-1}$) }   &       &    &    &   &     &     &    \\ 
Outflow velocity  & 6.0$^6$ & - & 7.5$^7$ & - & - &7.8$^7$ & 12$^9$ \\
{\em(km s$^{-1}$) }   &       &    &    &   &     &     &    \\ 
\hline
\end{tabular}
\vskip .2in

\small{{\bf Table References:} 1. SIMBAD database; 2. Gezari et al. 1999; 3.
Whitelock et al. 2000b; 4. Perryman et al. 1997 ({\it Hipparcos} catalog);  5.
Underhill 1984;  6. Winters et al. 2003; 7. Knapp et al. 1998; 8. Drake
et al. 1986; 9. Loup et al. 1993;  10. Percy et al. 2001.  \\

$^{11}$ Luminosity is estimated using the relation L = 4$\pi$d$_*$$^2$F$_{bol}$ .} 
\end{table*}

\begin{table*}[p]
\begin{center}
\caption{\bf Derived Infrared angular diameters }
\vskip 0.2in 
\begin{tabular}{ccccc}
\hline
\hline
 Star  &  Date &  Phase  & $\lambda$/$\Delta\lambda$ & Ang. Dia (UD)    \\
 &  &  &  ($\mu$m) & (mas)  \\
\hline 
U Ari & 19 Feb 2002 & 0.57 & 2.2/0.4 & 7.3$\pm$0.3 \\
 
      &             &      &                &     \\ 
Z Sco & 22 Mar 2003 & 0.26 & 2.2/0.4 & 3.8$\pm$1.0\\
       &             &      &                &     \\      
U Ori  & 13 Mar 2000 & 0.33 & 2.2/0.4 & 11.9$\pm$0.3 \\
       &             &      &         &   \\
SW Vir & 01 Jun 2001 & 0.56  & 2.2/0.4& 15.9$\pm$0.6\\ 
      &             &      &                &    \\ 
$\eta$ Gem & 08 Jan 2001 &   & 2.2/0.4 &12.7$\pm$0.3  \\
  	& 08 Jan 2001 &   & 3.8/0.6  & 12.7$\pm$1.0  \\ 
  	& 04 Mar 2001 &   & 2.2/0.4  & 12.8$\pm$0.3\\
	& 04 Mar 2001 &   & 3.8/0.6  & 12.8$\pm$2.0\\
       &             &      &                &  \\ 
$\mu$ Gem & 04 Mar 2001 &  & 2.2/0.4 & 13.7$\pm$0.5  \\
 	 & 04 Mar 2001 &  & 3.8/0.6 & 14.8$\pm$1.0 \\ 
      &             &      &                &      \\        
TV Gem & 14 Nov 2000 &  & 2.2/0.4  & 4.8$\pm$0.2 \\  
\hline
\end{tabular}
\end{center} 
\end{table*} 

\begin{table*}[p]
\caption{ \bf{Previous Angular size measurements from Optical to 
Near-IR}} 
\vskip 0.1in
 
\begin{tabular}{ccccccl}  
\hline
\hline
Star  & Date of Obs.    & Method$^a$ &  Phase    &  $\lambda$/$\Delta\lambda$  & Ang. Dia(UD)   
& Reference \\
 &     &   &   & ($\mu$m)  & (mas)  &  \\
\hline
U Ari & 03 Sep 1977 & LO   & 0.49 & 1.62/0.42 & 6.11 $\pm$0.34 &  
Ridgway et al. 1979 \\
(Mira)&  &    &  &   &   & \\
&   &     &  &  &  &   \\

U Ori  & 15 Oct 2000 & LBI & 0.88 & 2.20/0.40 & 15.59$\pm$0.06 & Mennesson et
al. 2002 \\
(Mira) &  15 Nov 2000 & LBI & 0.96 & 3.75/0.70 & 25.66$\pm$0.69 & Mennesson et
al. 2002 \\
       &  Oct 2000   & LBI & 0.83 & 2.20/0.1 &  10.6 & Perrin et al. 2004 \\
       &  Nov 2001   & LBI & 0.91 & 2.20/0.1 &  9.66$\pm$0.12 & Perrin et al.
       2004 \\
 &   &     &  &  &  &   \\        
SW Vir & 26 Jan 1981  & LO  & 0.02 &1.62/0.04 & 16.82 $\pm$0.34 &  
Ridgway et al. 1982 \\
(SRb) & 01 Sep 1981   & LO   & 0.48 &2.17/0.03 & 16.11$\pm$0.13 & 
Schmidtke et al. 1986 \\
& 29 Jun 1982  & LO  & 0.48 &2.28/0.40 & 16.77$\pm$0.23 & 
Schmidtke et al. 1986 \\
&29 Feb 2000   & LBI & 0.51 & 2.20/0.40 & 16.24$\pm$0.06 & 
Mennesson et al. 2002 \\
& 12 Mar 2000   & LBI & 0.56 & 3.75/0.70 & 22.88$\pm$0.33 & 
Mennesson et al. 2002 \\
 
 &   &  &    &  &  &   \\ 
$\eta$ Gem  & 1988-90 & LBI & &0.55/0.02 & 11.43$\pm$0.55 & Mozurkewich et al. 2003 \\
(SRa) & 1988-90 & LBI & &0.80/0.02 & 10.91$\pm$0.11 & Mozurkewich et al. 2003 \\ 
& 1993 &  LBI &  &0.712/0.012 & 11.75$\pm$0.27 &  
Quirrenbach et al. 1993 \\
& 1993 & LBI &    & 0.754/0.005 & 10.70$\pm$0.15 &  
Quirrenbach et al. 1993 \\
& 31 Mar 2001 & LO &   & 2.20/0.40  & 12.57$\pm$0.04 & Richichi et al. 2003 \\   
&   &  &     &  &  &   \\

$\mu$ Gem & 1988-90   & LBI   &  & 0.55/0.02 & 13.48$\pm$0.19 & Mozurkewich et al. 2003 \\
(Lb)  &  1988-90  & LBI   &  & 0.80/0.02 & 13.99$\pm$0.14 & Mozurkewich et al. 2003 \\ 
  & 1993 & LBI &  & 0.754/0.005 & 13.50$\pm$0.13 &
 Quirrenbach et al. 1993 \\
  & 1993 & LBI  & &0.712/0.012 & 13.97$\pm$0.28 &
 Quirrenbach et al. 1993 \\ 
 & 1987 & LBI   &  & 2.2/0.40 & 13.50$\pm$0.15 &  Di Benedetto et al. 1987 \\
 &   &  &    &  &  &   \\
TV Gem  & 15 Aug 1982 &  LO  &  &0.55 & 5.31$\pm$0.91 &  
Radick et al. 1984 \\
(SRc) & 30 Mar 1993   & LO   &  & 2.2/0.40 & 4.9$\pm$0.30 & 
Ragland et al. 1997 \\
& 03 Feb 1993   & LO   &   & 2.2/0.40 & 4.46$\pm$0.07 &
Richichi et al. 1998  \\
\hline  
\end{tabular} 
\vskip 0.1in
$^a${\small{LO: Lunar Occultation; LBI: Long Baseline Interferometry}} 
\end{table*}

\begin{table*}[p] 
 
\caption{\bf{Derived angular diameters, bolometric fluxes and effective temperatures}} 
\vskip 0.2in
\begin{tabular}{ccccccc}   
\hline
\hline
Star & Spt.Type  & m$_k$  & m$_v$  & UD (K-band) &  F$_{bol}$ $\times$10$^{8}$   &  T$_{eff}$   \\  
 &  &   &    & mas    & (erg cm$^{-2} $ s$^{-1}$)    &(K)   \\  
\hline

U Ari & M9 IIIe   & 1.26  & 14.50 & 7.3$\pm$0.3 &  45$\pm$5   &
2250$\pm$80  \\ 
           &       &       &               &                &     &    \\	  
Z Sco & M4/5IIIe  & 1.33  & 11.6  & 3.8$\pm$1.0 & 46$\pm$5 & 3120$\pm$420 \\
          &       &       &               &                &     &    \\	  
U Ori  & M8 IIIe & -0.88 & 10.89 & 11.9$\pm$0.3$^a$ & 336$\pm$33 & 2905$\pm$80 \\
           &       &       &               &                &     &    \\
SW Vir & M7 III & -1.74 & 7.90  & 15.9$\pm$0.6 & 735$\pm$110  & 3060$\pm$130
 \\
          &       &       &               & &   &    \\
$\eta$ Gem & M2.5 III & -1.49 & 3.70 & 12.8$\pm$0.3 
 
& 760$\pm$105 & 3450$\pm$125 \\
           &      &       &               &                &           &    \\ 
$\mu$ Gem & M3 III & -1.89 & 3.20  & 13.7$\pm$0.5 & 1140$\pm$170 &
3675$\pm$140   \\ 
          &       &       &                &  &   &   \\
TV Gem & M1 Ia & 1.16   & 6.83  & 4.8$\pm$0.2  &  153$\pm$15  & 3750$\pm$120  
 \\
\hline
\end{tabular}
\vskip 0.1in
$^a${\small{The angular diameter of U Ori is taken from  \citet{mon04}}}   
  
\end{table*} 

\begin{table*}[p]
\caption{\bf Linear radii and distances of Mira and SRs } 
\vskip 0.2in
{\small 
\begin{tabular}{cccccccc}
\hline
\hline  
 Star & K-band UD & Period & PL dist.  & Hip.
 dist.  & adopted dist. &  Lin. radii & Q-value  \\
& (mas) & (days)  & (pc)  & (pc) 
& (pc)   &   (R$_\odot$) & (for 1 M$_\odot$)   \\
\hline
U Ari & 7.3$\pm$0.3 &  371 & 776$\pm$155$^a$  & - & 776$\pm$155 & 610$\pm$125 &
0.025 \\ 
 
  &  &  &  &  &  &     \\   
Z Sco & 3.8$\pm$1.0 &  352 & 778$\pm$145$^a$  & 585$\pm$875$^b$ & 778$\pm$145 &
320$\pm$100& 0.062\\ 
   &  &  &  &  &  &     \\
U Ori & 11.9$\pm$0.3 & 372 & 306$\pm$61$^a$ & 505$\pm$420$^b$  & 306$\pm$61 &
391$\pm$78 & 0.048  \\ 
  &  &  &  &  &  &  &   \\
SW Vir & 15.9$\pm$0.6 &  150 & 98$\pm$19$^a$ & 143$\pm$24  & 143$\pm$24 &
244$\pm$42 & 0.039 \\
  &  &  &  &     &   & & \\ 
$\eta$ Gem & 12.8$\pm$0.3 & 233 & 150$\pm$30$^a$ & 107$\pm$22 & 107$\pm$22 &
146$\pm$30 & 0.130 \\
   &  & 20$^c$ &   &  & & & 0.012  \\
   &  &  &  &  &  &  &  \\
$\mu$ Gem & 13.7$\pm$0.5  & 27   &  -    &71$\pm$5    & 71$\pm$5 & 104$\pm$8 & 
0.025 \\
   &  &  &  &  &  &  &  \\
TV Gem & 4.8$\pm$0.2  & 182  &  -  & 1492$\pm$2340 & 1200$\pm$300$^d$ &
623$\pm$158 & 0.012 \\
   &  &  &  &  &  &  &  \\
\hline
\end{tabular}
\vskip 0.1in
$^a${\small \em From Period-Luminosity(PL) relation of Whitelock \& Feast
(2000b).} \\
$^b${\small \em Revised Hippaecos parallax from Knapp et al. (2003).} \\
$^c${\small \em This secondary period is determined from visual photometric
observations by Percy et al. (2001). } \\ 
$^d${\small\em Adopted from Underhill (1984). }}
\end{table*}


\begin{thebibliography}{}
\bibitem[Baize (1980)]{bai80} Baize, P. 1980, \aaps, 39, 83
\bibitem[Bedding et al. (1998)]{bed98} Bedding, T.R, Zijlstra, A.A., Jones, A. \& Foster, G. 1998, \mnras, 301, 1073
\bibitem[Bedding \& Zijlstra (1998)]{bed98a} Bedding, T.R \& Zijlstra, A.A. 1998,
\apjl, 506, L47
\bibitem[Bessell, Scholz \& Wood (1996)]{bes96} Bessell, M.S., Scholz, M. \& Wood, P.R.
1996, \aap, 307, 481
\bibitem[Bessell, Castelli \& Plez (1998)]{bes98} Bessell, M.S., Castelli, F. \& Plez, B. 1998, \aap, 333, 231
\bibitem[Catchpole et al. (1979)]{cat79} Catchpole, R.M., Robertson, B.S.C., Lloyds Evans, T.H.H, Feast,
M.W.,Glass, I.S. \& Carter, B.S. 1979, South Astronomical Astronomy Observatory Circular, 1, 61
\bibitem[Chandrasekhar (1999)]{chand99} Chandrasekhar, T. 1999, Bull. Astr. Soc. India, 27, 43
\bibitem[Chapman, Cohen \& Saikia (1991)]{chap91} Chapman, J.M., Cohen, R.J. \& Saikia, D.J. 1991, \mnras, 249, 227
\bibitem[Cho, Kaifu \& Ukita (1996)]{cho96} Cho, S.-H., Kaifu, N. \& Ukita, N. 1996, \aaps, 115, 117
\bibitem[Danchi et al. (1994)]{dan94} Danchi, W.C., Bester, M., Degiacomi, C.G., Greenhill, L.J. \& Townes, C.H.
1994, \aj, 107, 1469
\bibitem[Di Benedetto \& Rabbia (1987)]{diben87} Di Benedetto, G.P. \& Rabbia, Y. 1987, \aap, 188,114
\bibitem[Drake \& Linsky (1986)]{dra86} Drake, S.A.\& Linsky, J.L. 1986, \aj, 91, 602
\bibitem[Drake et al. (1991)]{dra91} Drake, S.A., Linsky, J.L., Judge, P.G. \& Elitzur, M. 1991, \aj, 101, 230
\bibitem[Feast  (1996)]{fea96} Feast, M.W. 1996, \mnras, 278, 11
\bibitem[Fedele et al. (2005)]{fedele05} Fedele, D., et al. 2005, \aap, 431, 1019    
\bibitem[Fox \&  Wood (1982)]{fox82} Fox, M.W. \&  Wood, P.R. 1982, \apj, 259, 198
\bibitem[Gezari, Pitts \& Schmitz (1999)]{gez99} Gezari, D. Y., Pitts, P. S. \& Schmitz, M.  1999, Catalog of Infrared
Observation, 5th Ed., NASA, Ref. Pub. 2225
\bibitem[Haniff, Scholz \& Tuthill (1995)]{han95} Haniff, C.A., Scholz, M. \& Tuthill, P.G. 1995, \mnras, 276, 640
\bibitem[Hinkle, Lebzelter \& Scharlach (1997)]{hin97} Hinkle, K.H., Lebzelter, T. \& Scharlach, W.W.G.  1997,\aj, 114, 2686
\bibitem[Hinkle \& Barnes (1979)]{hin79} Hinkle, K.H. \&  Barnes, T.G. 1979,\apj, 227, 923
\bibitem[Hofmann \& Scholz (1998)]{hof98} Hofmann, K, -H \& Scholz, M. 1998, \aap, 335, 637
\bibitem[Hofmann, Scholz \& Wood (1998)]{hof98a} Hofmann, K, -H, Scholz, M. \& Wood P.R. 1998, \aap, 339, 846
\bibitem[H$\ddot{o}$fner et al. (1998)]{hofner98} H$\ddot{o}$fner, S., Jorgensen, U.G., Loidl, R. \& Aringer, B. 1998,\aap, 340, 497
\bibitem[H$\ddot{o}$fner et al. (2003)]{hofner03} H$\ddot{o}$fner, S., Gaustschy-Loidl, R., Aringer, B. \& Jorgensen, U.G.  2003,\aap, 399, 589
\bibitem[Ireland et al. (2004a)]{ire04a} Ireland, M.J., Tuthill, P.G., Bedding, T.R., Robertson, J.G. \& Jacob, A.P. 2004a, \mnras, 350, 365
\bibitem[Ireland et al. (2004b)]{ire04b} Ireland, M.J., Scholz, M. \& Wood, P.R., 2004b, \mnras, 352, 318
\bibitem[Ireland et al. (2004c)]{ire04c} Ireland, M.J., Scholz, M., Tuthill, P.G. \& Wood, P.R., 2004c, \mnras, 355, 444
\bibitem[Jacob \& Scholz (2002)]{jacob02} Jacob, A.P. \& Scholz, M., 2002, \mnras, 336, 1377
\bibitem[Jura \& Kleinmann (1992)]{jura92} Jura, M. \& Kleinmann, S.G.  1992, \apjs, 79, 105
\bibitem[Keenan \& McNeil (1989)]{keen89} Keenan P.C. \& McNeil, R.C. 1989, \apjs, 71, 245
\bibitem[Kiss et al. (1999)]{kiss99} Kiss, L.L., Szatmary, K., Cadmus, Jr., R.R. \& Mathei, J.A.  1999, \aap, 346, 542  
\bibitem[Knapp et al. (1998)]{knapp98} Knapp, G. R., Young, K., Lee, E. \& Jorrisen, A. 1998, \apjs, 117, 209
\bibitem[Knapp et al. (2003)]{knapp03} Knapp, G. R., Pourbaix, D., Platais, I. \& Jorrisen, A. 2003, \aap, 403, 993
\bibitem[Kukarkin et al. (1969)]{kuk69} Kukarkin, B.V. et al., 1969, {\em General Catague of Variable Stars},
3rd ed. (Astronomical Council of the Academy of Sciences in the U.S.S.R, Moscow), vol.1
\bibitem[Labeyrie et al. (1977)]{lab77} Labeyrie, A., Koechlin, L., Bonneau, D., Blazit, A. \& Foy, R. 1977, \apj, 218, L75
\bibitem[Lebzelter \&  Hron (1999)]{leb99} Lebzelter, Th. \&  Hron, J.  1999, \aap, 351, 533
\bibitem[Loup et al. (1993)]{loup93} Loup, C., Foryeille, T., Omont, A. \& Paul, J.F.1993, \aaps, 99, 291
\bibitem[Mathei (2004)]{math04} Mathei, J. A., 2004, Observations from the AAVSO  International Database (private communication)
\bibitem[Matsurra et al. (2002)]{mat02} Matsurra, M., Yamamura, I., Cami, J. \& Murakami, H., 2002, \aap, 383, 972
\bibitem[Mennesson  et al. (2002)]{menn02} Mennesson, B., et al.  2002, \apj, 579, 446
\bibitem[Millan-Gabet et al. (2005)]{millan05} Millan-Gabet, R., et al. 2005, \apj, 620, 961  
\bibitem[Mondal et al. (1999)]{mon99} Mondal, S., Chandrasekhar, T., Ashok, N.M. \&  Kikani, P.K.  1999, Bull. Astr. Soc. India, 27, 335  
\bibitem[Mondal, Chandrasekhar \& Kikani (2002)]{mon02} Mondal, S., Chandrasekhar, T. \& Kikani, P.K., 2002 , Bull. Astr. Soc. India, 30, 811
\bibitem[Mondal \& Chandrasekhar (2004)]{mon04} Mondal, S. \& Chandrasekhar, T. 2004, \mnras, 348, 1332
\bibitem[Mozurkewich  et al. (2003)]{moz03} Mozurkewich, D. et al. 2003, \aj, 126, 2502
\bibitem[Nather \& McCants (1970)]{nath70} Nather, R. E. \& McCants, M. M. 1970, \aj, 75,963 
\bibitem[Ohnaka (2004)]{ohnaka04} Ohnaka, K., 2004, \aap, 424, 1011
\bibitem[Ostile \& Cox (1986)]{ost86} Ostile, D.A. \& Cox, A.N. 1986, \apj, 311, 864
\bibitem[Perrin et al. (1998)]{per98} Perrin, G.,  et al., 1998, \aap, 331,619
\bibitem[Perrin et al. (2004)]{per04} Perrin,  G. et al. 2004, \aap, 426, 279
\bibitem[Percy \& Parkes (1998)]{percy98} Percy, J. R. \& Parkes, M. 1998, \pasp, 110, 1431
\bibitem[Percy \& Wilson (2001)]{percy01} Percy, J. R. \& Wilson, B.J. 2001, \pasp, 113, 983
\bibitem[Perryman et al. (1997)]{perry97} Perryman, M.A.C. et al., 1997, The Hipparcos and Tyco Catalogs (ESA SP-1200)(Noordwijk: ESA)
\bibitem[Phillips et al. (1980)]{phil80} Phillips, J. P., Selby, M. J., Wade, R. \& Sanchez Magro, C. 1980, \mnras, 190, 337
\bibitem[Quirrenbach et al. (1993)]{qui93} Quirrenbach, A., Mozurkewich, D., Armstrong, J.T., Buscher, D.F. \& Hummel, C.A. 1993, \apj, 406, 215 
\bibitem[Radick, Henry \& Sherlin (1984)]{rad84} Radick, R.R., Henry, G.W \& Sherlin, J.M. 1984, \aj, 89, 151
\bibitem[Ragland, Chandrasekhar \& Ashok (1997)]{rag97} Ragland, S., Chandrasekhar, T. \& Ashok, N.M. 1997, \aap, 319, 260
\bibitem[Richichi et al. (1998)]{ric98} Richichi, A., Ragland, S., Stecklum, B. \& Leinert, Ch. 1998, \aap, 338, 527 
\bibitem[Richichi \& Calamai (2003)]{ric03} Richichi, A. \& Calamai, G. 2003, \aap, 399, 275
\bibitem[Ridgway et al. (1979)]{rid79} Ridgway, S.T., Wells, D.C., Joyce, R.R.\& Allen, R.G. 1979, \aj, 84, 247
\bibitem[Ridgway et al. (1982)]{rid82} Ridgway, S.T., Jacoby, G.H., Joyce, R.R. Siegel, J.M. \&  Wells, D.C. 1982, \aj, 87, 808
\bibitem[Schmidtke et al. (1986)]{sch86} Schmidtke, P.C., Africano, J.L., Jacoby, G.H., Joyce, R.R. \& Ridgway, S.T. 1986, \aj, 91, 961
\bibitem[Scholz \& Wood (2000)]{scholz00} Scholz, M \& Wood, P.R. 2000, \aap, 362, 1065
\bibitem[Scholz (2003)]{scholz03} Scholz, M. 2003,  in Astronomical Telescope and Instrumentation - Interferometry for Optical
Astronomy II, ed. W. A. Traub, Proceedings of the SPIE, 4838, 163
\bibitem[Schuller et al. (2004)]{Schuller04} Schuller, P., et al., 2004, \aap,
418, 151
\bibitem[Smith et al. (2002)]{smith02} Smith, B.J., Leisawitz, D., Castelaz, M.W. \& Luttermoser, D. 2002,
\aj, 123,948
\bibitem[Sloan  \& Price (1998)]{sloan98} Sloan, G.C. \& Price, D.S. 1998, \apjs, 119, 141
\bibitem[Tej, Lancon \& Scholz (2003a)]{tej03a} Tej, A., Lancon, A. \& Scholz, M.
2003a, \aap, 401, 347
\bibitem[Tej et al. (2003b)]{tej03b} Tej, A., Lancon, A., Scholz, M. \& Wood,
P.R. 2003b, \aap, 412, 481
\bibitem[Tej et al. (1999)]{tej99} Tej, A., Chandrasekhar, T., Ashok, N.M.,
Ragland, S., Richichi, A. \& Stecklum, B. 1999, \aj, 117, 1857
\bibitem[Thompson, Creech-Eakman \& van Belle (2002)]{thomp02} Thompson, R.R., Creech-Eakman  M.J. \& van Belle G.T., 2002, \apj, 577, 447
\bibitem[Tsuji et al. (1997)]{tsuji97} Tsuji, T., Ohnaka, K., Aoki, W.
\& Yamamura, I. 1997, \aap, 320, L1
\bibitem[Tsuji (2001)]{tsuji01} Tsuji, T. 2001, \aap, 376, L1
\bibitem[Underhill (1984)]{und84} Underhill, A.B.  1984, \pasp, 96, 305
\bibitem[van Belle (1999)]{van99} van Belle, G.T. 1999, \pasp, 111, 1515
\bibitem[van Belle, Thompson \& Creech-Eakman (2002)]{van02} van Belle G.T., Thompson, R.R. \& Creech-Eakman, M.J.  2002, \aj,
124, 1706
\bibitem[van Leeuwen et al. (1997)]{vanl97}  van Leeuwen, F., Feast, M. W., Whitelock, P.A. \& Yudin, B.  1997,
\mnras, 287, 955
\bibitem[Weiner, Hale \& Townes (2003a)]{weiner03a} Weiner, J., Hale, D.D.S. \&
Townes, C.H. 2003a, \apj, 588, 1064
\bibitem[Weine, Hale \& Townes (2003b)]{weiner03b} Weiner, J., Hale, D.D.S. \&
Townes, C.H. 2003b, \apj, 589, 976
\bibitem[Weiner (2004)]{weiner04} Weiner, J., 2004, \apjl, 611, L37
\bibitem[White \& Feierman (1987)]{white87} White, N.M. \& Feierman, B.H.  1987, \aj, 94, 751
\bibitem[Whitelock, Marang \& Feast (2000a)]{white00a} Whitelock, P.A., Marang,
F. \& Feast, M.W. 2000a, \mnras, 319, 728
\bibitem[Whitelock \& Feast (2000b)]{white00b} Whitelock, P.A. \& Feast, M.W.  2000b, \mnras, 319, 759
\bibitem[Winters et al. (2003)]{win03} Winters, J.M., Le Bertre, T., Jeong, K.S., Nyman, L, -A. \&
Epchtein, N.  2003, \aap, 409, 715 
\bibitem[Wittkowski et al. (2001)]{wit01} Wittkowski, M., Hammuel, C.A.,
Johnston, K.J., et al. 2001, \aap, 377, 981
\bibitem[Wittkowski et al. (2004)]{wit04} Wittkowski, M., Aufdenberg, J.P.,
Kervella, P., 2004, \aap, 413, 711
\bibitem[Woitke et al. (1999)]{woitke99} Woitke, P., Helling, Ch.,
Winters, J.M. \& Jeong, K.S. 1999, \aap, 348, L17
\bibitem[Wood (1990)]{wood90} Wood, P.R. 1990, in Miras to Planetary Nebulae : which path for
stellar evolution ? ed. M. O. Mennessier and A. Omont (Editions
Fronti$\grave{e}$res: France ), 67
\bibitem[Wood et al. (1999)]{wood99} Wood, P.R. et al., 1999, in IAU Sypm. No. 191, Asymptotic giant branch
stars, eds. T. Le Bertre, A. Lebre, and C. Waelkens (San Francisco: ASP), p151
\bibitem[Woodruff et al. (2004)]{wruff04} Woodruff, H.C., et al. 2004, \aap, 421, 703
\bibitem[Wyatt \& Cahn (1983)]{wyatt83} Wyatt, S.P. \& Cahn, J.H. 1983, \apj, 275, 225
\bibitem[Xiong, Deng \& Cheng (1998)]{xiong98} Xiong, D.R., Deng, L. \& Cheng,
Q.L. 1998, \apj, 499, 355
\bibitem[Yamamura, de Jong \& Cami (1999)]{yam99} Yamamura, I., de Jong,
T. \& Cami, J. 1999, \aap, 348, L55  
\bibitem[Young (1995)]{young95} Young, K. 1995, \apj, 445, 872
\end{thebibliography}
\end{document}